\documentclass[12pt,preprint,numberedappendix,twocolappendix]{emulateapj}

\newcommand\bibinc{n}		% set to y if bib pasted in .tex, set to n to use bibtex

\usepackage{subeqnarray}
\usepackage{amsmath}
\usepackage{hyperref}
\usepackage{ulem}
\usepackage{stmaryrd}

\hypersetup{colorlinks=true,linkcolor=black,citecolor=blue,urlcolor=black}

\usepackage{natbib}
\def\tt{\bf   }

\newcommand{\Eq}[1]{Equation\,(\ref{#1})}

\newcommand{\Sec}[1]{Section~\ref{#1}}

\newcommand{\Fig}[1]{Figure~\ref{#1}}

\newcommand {\twostr} {{\ttfamily TWOSTR}}
\newcommand {\disort} {{\ttfamily DISORT}}
\newcommand {\mitgcm} {{\ttfamily MITgcm}}
\newcommand {\sparc} {{\ttfamily SPARC}}

\setcitestyle{citesep={,}}
\begin{document}

\slugcomment{Accepted at ApJ}

\shorttitle{Dayside-nightside temperature differences of hot Jupiters: comparison with observations}
\shortauthors{Komacek, Showman, \& Tan}

\title{Atmospheric Circulation of Hot Jupiters: Dayside-Nightside Temperature Differences. II. Comparison with Observations}
\author{Thaddeus D. Komacek$^1$, Adam P. Showman$^1$, and Xianyu Tan$^1$} \affil{$^1$Lunar and Planetary Laboratory, Department of Planetary Sciences,
 University of Arizona, Tucson, AZ, 85721 \\
\url{tkomacek@lpl.arizona.edu}} 
\begin{abstract}
The full-phase infrared light curves of low-eccentricity hot Jupiters show a trend of increasing fractional dayside-nightside brightness temperature difference with increasing incident stellar flux, both averaged across the infrared and in each individual wavelength band. The analytic theory of \cite{Komacek:2015} shows that this trend is due to the decreasing ability with increasing incident stellar flux of waves to propagate from day to night and erase temperature differences. %, due both to radiative damping and frictional drag. 
Here, we compare the predictions of this theory to observations, showing that it explains well the shape of the trend of increasing dayside-nightside temperature difference with increasing equilibrium temperature. Applied to individual planets, the theory matches well with observations at high equilibrium temperatures but, for a fixed photosphere pressure of $100 \ \mathrm{mbar}$, systematically under-predicts the dayside-nightside brightness temperature differences at equilibrium temperatures less than $2000 \ \mathrm{K}$. We interpret this as due to as the effects of a process that moves the infrared photospheres of these cooler hot Jupiters to lower pressures.
%We interpret this as likely due to as the effects of clouds muting the infrared emission from the nightside of the planet. 
We also utilize general circulation modeling with double-grey radiative transfer to explore how the circulation changes with equilibrium temperature and drag strengths. As expected from our theory, the dayside-nightside temperature differences from our numerical simulations increase with increasing incident stellar flux and drag strengths. We calculate model phase curves using our general circulation models, from which we compare the broadband infrared offset from the substellar point and dayside-nightside brightness temperature differences against observations, finding that strong drag or additional effects (e.g. clouds and/or supersolar metallicities) are necessary to explain many observed phase curves. 
%Doing so, we find that disequilibrium chemical effects are likely necessary to explain the systematically large $3.6 \ \mu \mathrm{m}$ phase curve amplitudes observed to date. 
\end{abstract}
\section{Introduction}
\label{sec:intro}
Gas giant planets with very small semi-major axes were the first class of exoplanet to be observed in radial velocity and transit \citep{Mayor:1995,Charbonneau_2000,Henry:2000} and, due to their size and hot atmospheres, are the best-characterized type of transiting exoplanet (for recent reviews, see \citealp{Heng:2014b,Crossfield:2015}). These ``hot Jupiters'' are expected to have strong eastward winds at the equator comparable to or greater than the speed of sound in the medium \citep{Cooper:2005,Menou:2009,Thrastarson:2010,Showmanetal_2009,Heng:2011,perna_2012,Rauscher_2012,Dobbs-Dixon:2013,Mayne:2014}, due to an equatorial jet driven by the difference between the large incident stellar flux onto the dayside and lack of incident flux on the nightside of these presumably tidally locked objects \citep{showman_2002,Showman_Polvani_2011}. They have been observed in many cases to have a corresponding shift of the hottest point of the planet eastward of the substellar point (e.g. \citealp{Knutson_2007}), as predicted by general circulation models (GCMs), e.g. \cite{showman_2002}. More recently, observations of the Doppler shifts of spectral lines have allowed direct measurement of the rotation and winds of hot Jupiters \citep{Snellen:2010,Brogi:2015,Louden:2015}, with all observations finding fast wind speeds of a few $\mathrm{km} \ \mathrm{s}^{-1}$. 
\begin{figure}
	\centering
	\includegraphics[width=0.5\textwidth]{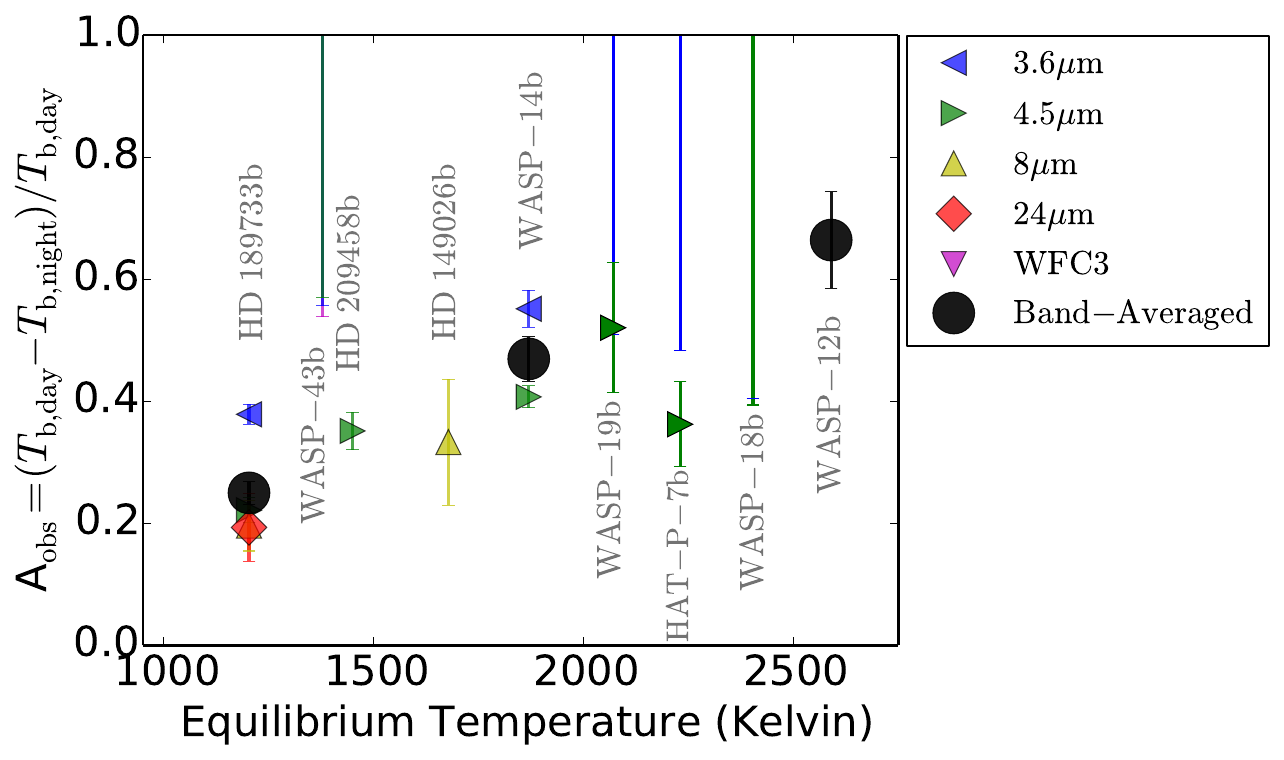}
	\caption{Observations of the fractional dayside-nightside brightness temperature difference A$_{\mathrm{obs}}$ for low-eccentricity transiting hot Jupiters in different infrared wavelength bands, plotted against global-average equilibrium temperature. Here we define the global-average equilibrium temperature $T_{\mathrm{eq}} = [F_{\star}/(4\sigma)]^{1/4}$, where $F_{\star}$ is the incoming stellar flux and $\sigma$ the Stefan-Boltzmann constant. For planets with observations in multiple wavelengths, a band-averaged value is computed as in \cite{Komacek:2015}. Lines without a point display lower limits on A$_{\mathrm{obs}}$, calculated using upper limits of observed flux. There is a general trend of increasing A$_{\mathrm{obs}}$ with increasing $T_{\mathrm{eq}}$, for both the band-averaged values and in each individual wavelength band. Observational data is taken from \cite{Knutson_2007,Knutson:2009a,Knutson:2009,Nymeyer:2011,Cowan:2012,Crossfield:2012,Knutson:2012,Maxted:2013,Stevenson:2014,Zellem:2014,Wong:2015a,Wong:2015,Stevenson2016}.}
	\label{fig:Aobs}
\end{figure} \\
\indent Although hot Jupiters have strong winds, the infrared phase curves of these planets often have large amplitudes, indicating that the pressures probed on the nightside are much colder those on the dayside. \Fig{fig:Aobs} shows the fractional amplitude, $A_{\mathrm{obs}}$, of the dayside-nightside brightness temperature difference, plotted against full-redistribution equilibrium temperature for observations in different wavelength bands of low-eccentricity transiting hot Jupiters. As shown by \cite{Cowan_2011,Perez-Becker:2013fv,Schwartz:2015}, and \cite{Komacek:2015}, there is a general trend, in a band-averaged sense, of increasing $A_{\mathrm{obs}}$ with increasing equilibrium temperature. \Fig{fig:Aobs} shown here is an update of the figure from \cite{Komacek:2015} that now includes how $A_{\mathrm{obs}}$ varies with with wavelength for each given planet. As a result, we also find that $A_{\mathrm{obs}}$ increases with equilibrium temperature in each individual wavelength band. \\          
\indent The trend shown in \Fig{fig:Aobs}, which indicates that the fractional dayside-nightside temperature differences in hot Jupiter atmospheres increase with increasing equilibrium temperature, has recently been analyzed theoretically by \cite{Perez-Becker:2013fv} using a one-layer fluid dynamical approach and by \cite{Komacek:2015} in three dimensions. These works showed that the trend in \Fig{fig:Aobs} can be interpreted as due to the decreasing efficacy of wave adjustment processes with increasing incident stellar flux. The large day-to-night stellar heating contrast in hot Jupiter atmospheres drives the existence of standing, large-scale equatorial wave modes \citep{Showman:2010,Showman_Polvani_2011,Tsai:2014}. These equatorial wave modes induce horizontal convergence/divergence, which leads to vertical motion that moves isentropes vertically. Although these wave modes are standing, the extent to which they extend longitudinally---and therefore modify the thermal structure between day and night---can be interpreted in terms of the ability (or inability) of the corresponding freely propagating modes to propagate zonally without being damped (see \citealp{Showman_Polvani_2011} and \citealp{Perez-Becker:2013fv} for further discussion of the wave interpretation of this theory). If these waves are not damped, and are able to propagate from day to night, this process then tends to promote a final state with flat isentropes \citep{Showman_2013_terrestrial_review}. A similar wave adjustment process occurs in equatorial regions of Earth's atmosphere \citep{Polvani:2001,Sobel:2001,Bretherton:2003}, and may occur in tidally locked terrestrial planet atmospheres \citep{Showman_2013_terrestrial_review,Koll:2014,Wordsworth:2014,Koll:2016}, which leads to small horizontal temperature differences and the often-invoked ``weak temperature gradient'' limit wherein horizontal temperature gradients are assumed to vanish.     \\
%In Paper I, \cite{Komacek:2015} showed that the weak temperature gradient limit is relevant in hot Jupiter atmospheres on the global scale. Using this limit, 
\indent In Paper I, \cite{Komacek:2015} extended the work of \cite{Perez-Becker:2013fv} to three dimensions by obtaining a fully analytic prediction of the dayside-nightside temperature contrast and characteristic wind speeds in an idealized model over a wide range of equilibrium temperature, frictional drag strengths, rotation rates, and atmospheric compositions. They compared this analytic theory to GCM calculations with a simplified Newtonian cooling scheme and found that the theory applies well for atmospheric conditions relevant to hot Jupiters. As in \cite{Perez-Becker:2013fv}, they showed that the combined effects of increasing relative efficacy of radiative cooling and increasing drag strength can explain the increase in dayside-nightside temperature contrast with increasing equilibrium temperature, as both mechanisms damp wave adjustment processes. This drag is likely due to either Lorentz forces in an ionized atmosphere threaded by a dipolar magnetic field \citep{Perna_2010_1,batygin_2013,Rauscher_2013,Rogers:2014,Rogers:2020}, or small-scale instabilities \citep{Li:2010,Youdin_2010,Fromang:2016}, including shocks, which largely are expected to occur on the dayside of the planet \citep{Heng:2012a,Fromang:2016}. However, \cite{Komacek:2015} found that drag only affects dayside-nightside temperature differences if it occurs on a characteristic timescale much shorter than the rotation period of the planet. As a result, the radiative damping of equatorial wave propagation can alone explain the trend of increasing $A_{\mathrm{obs}}$ with increasing equilibrium temperature, with drag likely only playing a second-order role. \\
\indent In this paper, we first compare the analytic theory from \cite{Komacek:2015} directly to observations, using the extension of the theory by \cite{Zhang:2016} that incorporates all possible dynamical regimes (\Sec{sec:comparisontoobs}). We then utilize three-dimensional numerical models of the atmospheric circulation with double-grey radiative transfer to understand how the relevant dynamics governing the day-night temperature contrast in hot Jupiter atmospheres varies with parameters of the circulation. We describe the numerical setup of these simulations in \Sec{sec:numsetup}, including a detailed description of the radiative transfer scheme, as this is the first application of \twostr \ (the two-stream mode of \disort) to hot Jupiter atmospheres.
%In order to confirm that the qualitative predictions of this theory (which uses a simplified linear heating/cooling scheme) apply to observations, we use a suite of numerical simulations with simplified double-grey radiative transfer. These simulations use an opacity profile that is not dependent on temperature, and as a result they can only confirm theoretical predictions qualitatively, not quantitatively\footnote{However, we performed detailed quantitative comparison between analytic predictions and numerical simulations using Newtonian cooling in \cite{Komacek:2015}.}. 
Using these simulations, we explore how the atmospheric circulation of hot Jupiters changes with varying equilibrium temperature from $500 - 3000 \ \mathrm{K}$ and drag timescales from $10^3 - \infty \ \mathrm{s}$, keeping the rotation rate fixed (\Sec{sec:trendsnumerical}). From these numerical simulations, we calculate in \Sec{sec:phasecurvecalc} how the simulated phase curves vary with equilibrium temperature and drag strength, and compare our simulated phase curve amplitudes to observations in \Sec{sec:bandresults}.
%should change with wavelength (\Sec{sec:bandresults}) to aid in the explanation of the observed $A_{\mathrm{obs}}$ in different wavelength bands. 
Additionally, we numerically explore the effects of rotation rate on the atmospheric circulation, varying the equilibrium temperature and rotation rate consistently to explore the effect rotation rate has on the dynamics of our main grid of simulations (\Sec{sec:consistentomega}). Lastly, we analyze our numerical infrared phase offsets, comparing them to both observations and the analytic theory of \cite{Zhang:2016} that allows for estimation of phase offsets (\Sec{sec:phaseoffsets}). We discuss our results and potential future avenues for theoretical work in \Sec{sec:discussion}, and express conclusions in \Sec{sec:conclusions}.   
\section{Dayside-nightside temperature differences: comparison of analytic theory with observations}
\label{sec:comparisontoobs}
\subsection{Trends with varying stellar irradiation and drag strengths}
\subsubsection{Theory}
\label{sec:theoryderiv}
In \cite{Komacek:2015}, we derived approximate steady-state analytic solutions for dayside-nightside temperature differences relative to those in radiative equilibrium.
%, see Section 4 of \cite{Komacek:2015}. 
This theory was derived using the weak temperature gradient limit, with vertical entropy advection globally balancing radiative heating/cooling (parameterized by a Newtonian cooling scheme), as \cite{Komacek:2015} showed that the vertical entropy advection term is larger (in a globally-averaged sense) than the horizontal entropy advection term at photospheric pressure levels ($\gtrsim 10 \ \mathrm{mbar}$) in hot Jupiter atmospheres. The solutions in \cite{Komacek:2015} were written for all possible balances in the momentum equation, where the day-night pressure-gradient force is balanced near the equator by either advection or drag and at high latitudes by either the Coriolis force, advection, or drag. \\
\indent Applying the theory of \cite{Komacek:2015} to explain trends in day-night temperature differences from their GCMs with varying mean molecular weight and heat capacity, \cite{Zhang:2016} derived a uniform expression for the dayside-nightside temperature difference that incorporates all dynamical terms (see their Appendix A). In this work, we use their complete solution for dayside-nightside temperature differences in order to compare to infrared observations of hot Jupiter phase curves. Their solution for approximate dayside-nightside temperature differences (relative to that in radiative equilibrium) is
\begin{equation}
\label{eq:deltaTsoln}
\frac{\Delta T}{\Delta T_{\mathrm{eq}}} \sim 1 - \frac{2}{\alpha + \sqrt{\alpha^2 + 4 \gamma^2}} \mathrm{,}
\end{equation} 
where 
\begin{equation}
\label{eq:alpha}
\alpha = 1 + \frac{\left(\Omega + \tau^{-1}_{\mathrm{drag}}\right)\tau^2_{\mathrm{wave}}}{\tau_{\mathrm{rad}} \Delta \mathrm{ln} p} \mathrm{,}
\end{equation}
and
\begin{equation}
\label{eq:gamma}
\gamma = \frac{\tau^2_{\mathrm{wave}}}{\tau_{\mathrm{rad}}\tau_{\mathrm{adv,eq}}\Delta \mathrm{ln} p} \mathrm{.}
\end{equation}
In Equations (\ref{eq:deltaTsoln})-(\ref{eq:gamma}) above, all variables retain their meaning from \cite{Komacek:2015} and \cite{Zhang:2016}\footnote{To summarize: the day-night temperature difference is $\Delta T$, the day-night temperature difference in radiative equilibrium is $\Delta T_{\mathrm{eq}}$, the rotation rate is $\Omega$, the characteristic drag timescale is $\tau_{\mathrm{drag}}$, the (Kelvin) wave propagation timescale across a hemisphere is $\tau_{\mathrm{wave}} = a/(NH)$ with $a$ the radius of the planet, $N$ is the Brunt-V{\"a}is{\"a}l{\"a} frequency (evaluated in the isothermal limit) and $H=RT/g$ is scale height, where $R$ is the specific gas constant, $T$ is temperature and $g$ is gravitational acceleration, and the radiative timescale is $\tau_{\mathrm{rad}}$. $\Delta \mathrm{ln}p$ is the number of scale heights from the pressure of interest to the level at which the dayside-nightside temperature difference goes to zero, taken to be $10 \ \mathrm{bars}$. $\tau_{\mathrm{adv,eq}} = a\sqrt{2/\left(R \Delta T_{\mathrm{eq}} \Delta \mathrm{ln}p\right)}$ is an advective timescale of the cyclostrophic wind induced by the day-night temperature difference in radiative equilibrium (i.e., one can rewrite $\tau_{\mathrm{adv,eq}} = a/U_{\mathrm{eq}}$.)}. \\ 
\indent For the comparison with observed results below, we calculate dayside-nightside temperatures at a pressure of $100 \ \mathrm{mbar}$, similar to the level of the infrared photosphere for a typical hot Jupiter. We take a planetary radius $a = 9.43 \times 10^7 \ \mathrm{m}$, specific gas constant $R = 3700 \ \mathrm{J} \ \mathrm{kg}^{-1} \mathrm{K}^{-1}$, specific heat $c_p = 1.3 \times 10^4 \mathrm{J} \ \mathrm{kg}^{-1} \mathrm{K}^{-1}$, and gravity $g = 9.36 \ \mathrm{m} \ \mathrm{s}^{-2}$. We assume that in radiative equilibrium the nightside temperature is small relative to the dayside temperature, and hence that $\Delta T_{\mathrm{eq}} = T_{\mathrm{eq}}$. 
%{\bf Note that such a radiative equilibrium cannot be achieved when including dynamics (e.g. \citealp{showman_2002,Rauscher:2010}), so our assumption of a base state in radiative equilibrium is only  valid if there is no atmospheric circulation in this state.} 
Additionally, we scale $\tau_{\mathrm{rad}}$ from \cite{Komacek:2015} using the power-law relationship with temperature from \cite{showman_2002} and \cite{Ginzburg:2015a}, setting
\begin{equation}
\label{eq:taurad}
\tau_{\mathrm{rad}}(p,T_{\mathrm{eq}}) = 10^5 \ \mathrm{s}\left(\frac{p}{100 \ \mathrm{mbar}}\right)\left(\frac{1800 \ \mathrm{K}}{T_{\mathrm{eq}}}\right)^3 \mathrm{.}
%\tau_{\mathrm{rad}} = 10^5 \mathrm{s} \left(\frac{1800 \ \mathrm{K}}{T_{\mathrm{eq}}}\right)^3 
\end{equation}
\subsubsection{Comparison to observations}
\begin{figure}
	\centering
	\includegraphics[width=0.5\textwidth]{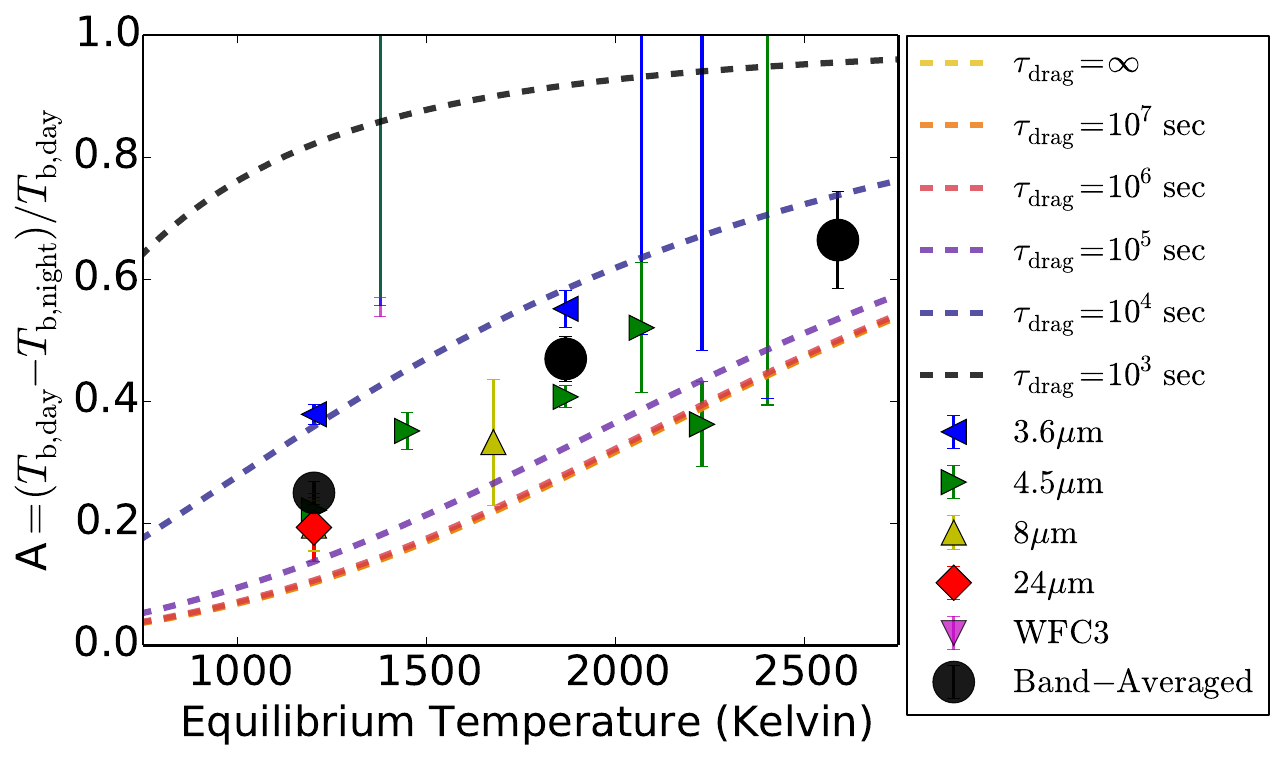}
	\caption{Comparison between theoretical analytic predictions for fractional dayside-nightside temperature differences, plotted for potential frictional drag timescales from $10^3 \ \mathrm{s} - \infty$, and observations. The theoretical predictions use the solutions that combine all dynamical regimes of the theory from \cite{Komacek:2015}, see Appendix A of \cite{Zhang:2016}. The theoretical predictions are for the $100 \ \mathrm{mbar}$ pressure level, approximately equal to the pressure of the infrared photosphere for a typical hot Jupiter. A rotation period of $3.5 \ \mathrm{days}$ is assumed, along with a scaling of $\tau_{\mathrm{rad}} \propto T_{\mathrm{eq}}^{-3}$ as in \cite{showman_2002} and \cite{Ginzburg:2015a}. Observations are the same as those shown in Figure \ref{fig:Aobs}. The theoretical predictions capture well the general slope of the trend of increasing A with increasing $T_{\mathrm{eq}}$. Note that the theoretical predictions when $\tau_{\mathrm{drag}} \geq 10^6 \ \mathrm{s}$ are the same, as $\tau_{\mathrm{drag}} \gg \Omega^{-1}$ and \Eq{eq:deltaTsoln} becomes independent of $\tau_{\mathrm{drag}}$.}
	\label{fig:A_theory}
\end{figure}
\begin{figure}
	\centering
	\includegraphics[width=0.5\textwidth]{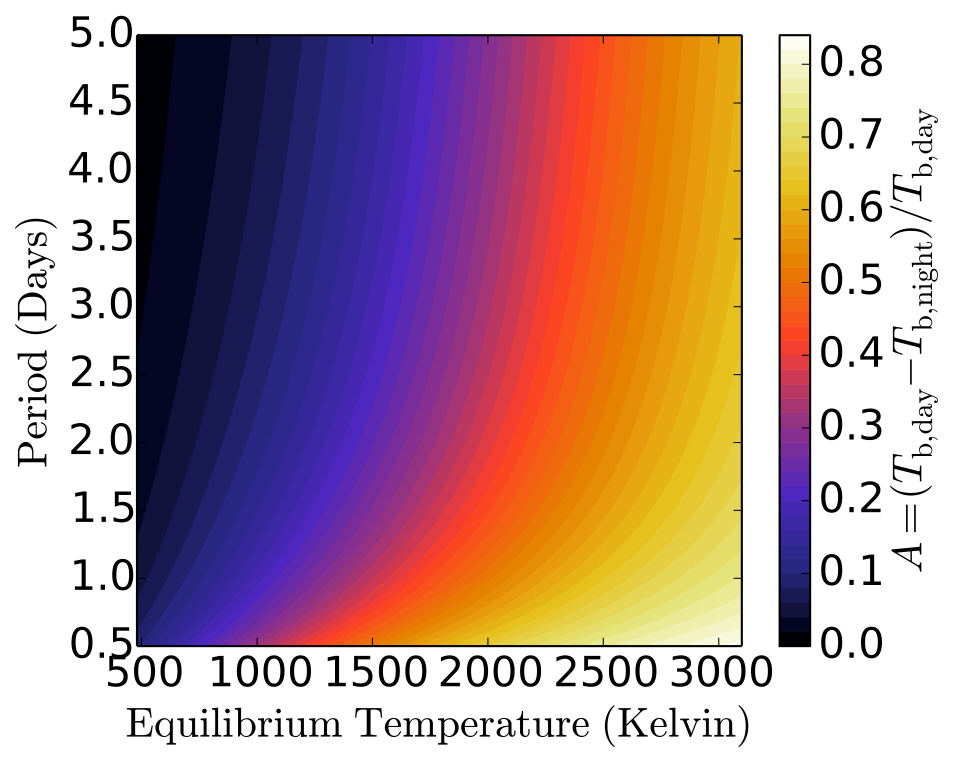}
	\caption{Theoretical predictions for the fractional dayside-nightside temperature difference $A$ as a function of rotation period and equlibrium temperature from \Eq{eq:deltaTsoln}, assuming $\tau_{\mathrm{drag}} = \infty$ (i.e. no drag in the observable atmosphere). As in \Fig{fig:A_theory}, the theoretical predictions are for the $100 \ \mathrm{mbar}$ pressure level. Though the incident stellar flux is the key parameter controlling dayside-nightside temperature differences, the rotation period can greatly affect $A$ if it is short. This is because at short rotation periods and hence fast rotation rates the global-average Coriolis force can become greater than advective forces, increasing the dayside-nightside pressure gradient that can be supported by the circulation.}
	\label{fig:A_teq_omega}
\end{figure}

\Fig{fig:A_theory} shows theoretical predictions for the fractional dayside-nightside temperature difference $A$ with a fixed rotation period of $3.5 \ \mathrm{days}$ and $\tau_{\mathrm{drag}}$ varying from $10^3 - \infty \ \mathrm{s}$, compared to the set of observed fractional dayside-nightside brightness temperature differences $A_{\mathrm{obs}}$. Here we take a long rotation period to effectively bracket the lower limit of expected dayside-nightside temperature differences, as a decreased rotation period increases the resulting dayside-nightside temperature differences (see \Fig{fig:A_teq_omega} and related discussion below). The theoretical predictions at low $\tau_{\mathrm{drag}}$ match well the trend of steadily increasing $A_{\mathrm{obs}}$ with increasing equilibrium temperature. However, note that the predictions for $\tau_{\mathrm{drag}} = 10^5 - \infty \ \mathrm{s}$ regularly under-predict the individual and band-averaged values of $A_{\mathrm{obs}}$\footnote{The solutions for $\tau_{\mathrm{drag}} = 10^6 - \infty \ \mathrm{s}$ overlap because $\tau_{\mathrm{drag}} >> \Omega^{-1}$ and therefore \Eq{eq:alpha} becomes $\alpha \approx 1 + \Omega \tau^2_{\mathrm{wave}}/(\tau_{\mathrm{rad}} \Delta \mathrm{ln}p)$, independent of $\tau_{\mathrm{drag}}$.}. 
\\ \indent There are three ways to increase dayside-nightside temperature differences relative to the prediction for the case with $\tau_{\mathrm{drag}} = \infty$. First, as seen in \Fig{fig:A_theory}, atmospheric drag can be strong, with a characteristic timescale shorter than $10^5 \ \mathrm{s}$. In the context of Lorentz ``drag,'' the variations from planet-to-planet could be caused by changing values of internal magnetic field strength or varying atmospheric composition and hence thermal ionization fraction. Second, we are here considering a relatively long rotation period (relevant for HD 209458b), and considering shorter rotation periods will cause the predicted value of $A$ to increase. This is because if the Coriolis force is stronger (due to a larger $\Omega$), it can support a larger day-to-night pressure gradient (which is related to the day-night temperature contrast) in steady-state. This can be seen in \Fig{fig:A_teq_omega}, which shows how the theoretical prediction of dayside-nightside temperature differences varies with the rotation period and incident stellar flux, assuming no atmospheric drag. Though incident stellar flux is the dominant factor controlling the dayside-nightside temperature contrast, planets with very short rotation periods (where the Coriolis force is stronger than the advective terms) have significantly larger day-night temperature differences. We will include the effects of rotation rate to directly compare with observations in \Sec{sec:indivpred}. Lastly, as will be discussed further in \Sec{sec:indivpred}, any process that shifts the optical-depth-unity surface to lower pressures can explain an increase in $A$. One example of this is supersolar metallicities, which have been shown to increase the day-night temperature difference in GCMs (e.g. \citealp{Showmanetal_2009,Kataria:2014,Wong:2015a}). Another possibility is high-altitude clouds, which are expected to form largely on the nightside and western limb \citep{Oreshenko:2016,Lee:2016,Parmentier:2015}, where it is coldest. As a result, clouds could increase the amplitude of the phase curve and thereby raise our theoretical predictions for $A$ closer to observed values. 
%Lastly, one can explain an increase in $A$ by the effects of clouds diminishing the outgoing infrared flux. Given that clouds are expected to form largely on the nightside and western limb \citep{Lee:2016,Parmentier:2015}, where it is coldest, it is likely that the effects of clouds would increase the amplitude of the phase curve and thereby raise our theoretical predictions for $A$ closer to observed values. 
\subsection{Predictions for individual planets}
\label{sec:indivpred}
\begin{figure}
	\centering
	\includegraphics[width=0.5\textwidth]{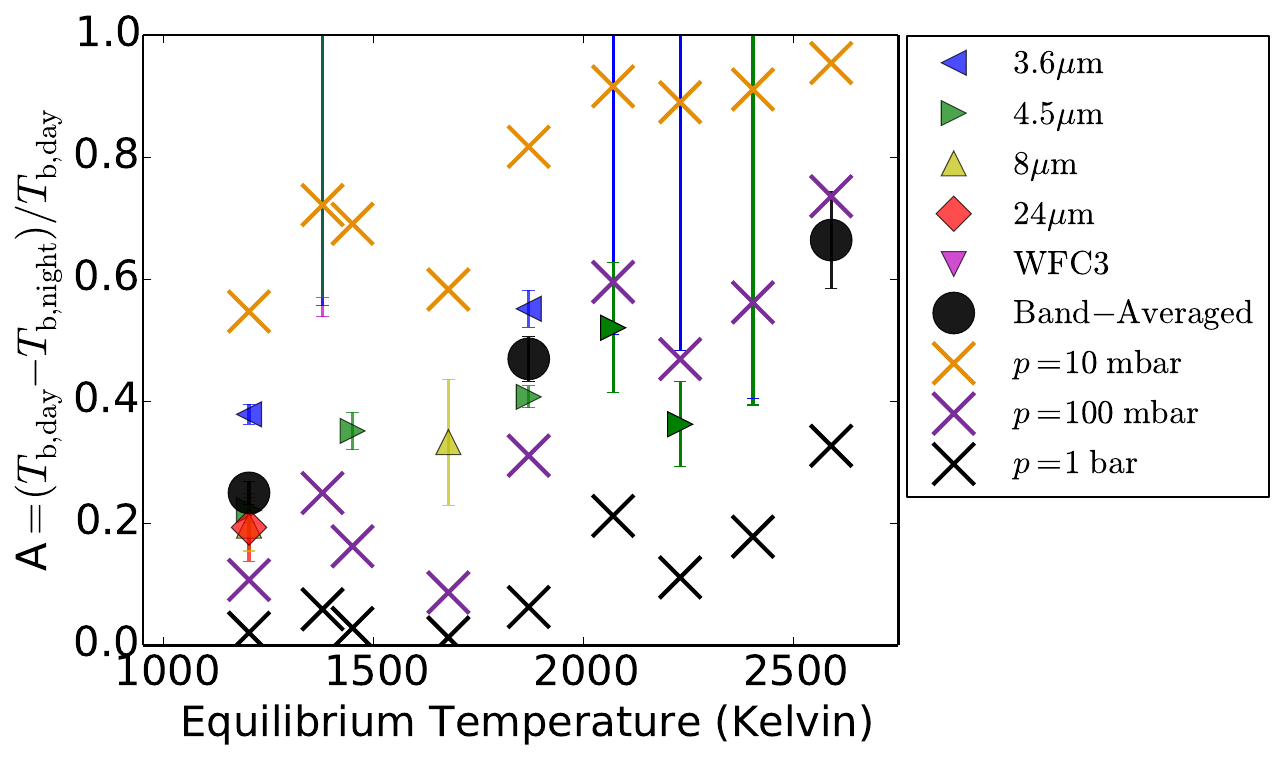}
	\caption{Comparison between theoretical predictions (at pressures of of $10$, $100$, and $1000 \ \mathrm{mbar}$) and observations of fractional dayside-nightside brightness temperature differences for each individual low-eccentricity transiting planet with full-phase infrared observations. The theoretical predictions use a rotation period for each planet that is equal to the orbital period, and assume no atmospheric frictional drag ($\tau_{\mathrm{drag}} = \infty$). As a result, these theoretical predictions are lower limits on $A$, as very strong drag would increase dayside-nightside temperature differences. Observations are the same as those shown in Figure \ref{fig:Aobs}. The theory at $100 \ \mathrm{mbar}$ pressure (near the expected infrared photosphere of hot Jupiters) captures the observed fractional dayside-nightside temperature differences well, especially at high $T_{\mathrm{eq}}$. This shows that strong frictional drag is not necessary to explain the large dayside-nightside temperature differences at high levels of incident stellar flux. Notably, this is where Lorentz forces are strongest due to the increase of the electrical conductivity of the atmosphere with increasing temperature. There is a large change in $A$ with an order of magnitude change in the photosphere pressure, potentially allowing explanation of the high observed day-night temperature contrast for planets such as WASP-43b.}
	\label{fig:A_theory_individual}
\end{figure}
Using the theory from \cite{Komacek:2015} and \cite{Zhang:2016} we can make predictions of $A$ for individual planets, assuming that they are tidally locked and hence that their orbital period is equal to their rotation period and choosing a drag strength. In \Fig{fig:A_theory_individual}, we plot theoretical predictions of $A$ for all planets from \Fig{fig:Aobs} along with their corresponding observations, setting $\tau_{\mathrm{drag}} = \infty$. We choose $\tau_{\mathrm{drag}} = \infty$ for these comparisons because this gives lower limits on the resulting predictions for fractional dayside-nightside temperature differences. The theoretical predictions at $100 \ \mathrm{mbar}$ (near the expected infrared photosphere of a typical hot Jupiter) match well with $A_{\mathrm{obs}}$ in the high-equilibrium temperature regime ($T_{\mathrm{eq}} \gtrsim 2000 \ \mathrm{K}$), and under-predict the values for all planets with $T_{\mathrm{eq}} \lesssim 2000 \ \mathrm{K}$. However, there is a strong dependence of the theoretical predictions for $A$ on pressure, with fractional day-night temperature differences near unity at pressures $\lesssim 10 \ \mathrm{mbar}$ and near zero at pressures $\gtrsim 1 \ \mathrm{bar}$. As a result, changing photosphere pressures between $10 - 100 \ \mathrm{mbar}$ in our theory brackets the range of fractional dayside-nightside temperature differences for all observed planets. Note that the short rotation period of $\approx 0.81 \ \mathrm{days}$ for WASP-43b does lead to a prediction of significantly larger day-night temperature contrast than for a slower rotating planet (e.g. HD 209458b). However, this increase in predicted day-night temperature difference due to the short rotation period is not by itself enough to provide a full quantitative explanation of the observed day-night temperature contrast of WASP-43b. \\
%, {\bf with either a low photosphere pressure or strong drag required to explain the observation}. \\
\indent Given that we are using a value of $\Omega$ consistent with the expected rotation rate of each planet, there are two possible explanations for the under-prediction of observed dayside-nightside temperatures by our theory assuming a photosphere pressure of $100 \ \mathrm{mbar}$ at low $T_{\mathrm{eq}}$: strong drag affecting the ability of the circulation to erase dayside-nightside temperature differences or photosphere pressures being $\lesssim 100 \ \mathrm{mbar}$. For the case of WASP-43b, the photosphere pressure must be lower than $25 \ \mathrm{mbar}$ to explain the phase curve amplitude without invoking strong atmospheric drag. Given that Lorentz forces are expected to be weak at low $T_{\mathrm{eq}} \lesssim 1400 \ \mathrm{K}$ \citep{Rogers:2014}, it is unlikely that magnetic effects alone can explain the discrepancy between theoretical predictions and observations at low $T_{\mathrm{eq}}$. \\
\indent A decreased photosphere pressure would be related to the presence of the large column abundance of infrared absorbers, which in this case could be cloud coverage muting the outgoing infrared flux. Nightside clouds would naturally tend to be more prevalent in the atmospheres of cooler hot Jupiters than those in hotter hot Jupiters. This is simply because cooler temperatures allow a greater variety of chemical species to condense. This could help explain why we are under-predicting the value of the dayside-nightside temperature difference primarily at the lowest equilibrium temperatures, while matching the observed dayside-nightside temperature differences at higher equilibrium temperatures. Specifically, clouds are expected to be present with an effective cloud coverage near unity on the nightside regardless of $T_{\mathrm{eq}}$, but the aerosol particle size is expected to increase with $T_{\mathrm{eq}}$ \citep{Spiegel:2009,Heng:2013,Parmentier:2015}. Using the difference in transit radii between the line center and wing of potassium and sodium lines, \cite{Heng:2016} showed that hot Jupiters with $T_{\mathrm{eq}} \lesssim 1300 \ \mathrm{K}$ are cloudier than hotter planets.
%Though Lorentz forces may contribute to the relatively large observed dayside-nightside temperature differences at low $T_{\mathrm{eq}}$,\cite{}
Additionally, observations have shown that clouds are nearly ubiquitous for cool planets with $T_{\mathrm{eq}} \lesssim 2000 \ \mathrm{K}$ \citep{Demory_2013,Esteves:2015,Sing:2015a}. Another possibility is supersolar metallicities, which increase the opacity of the atmosphere and thereby adjust the photosphere upward. However, it is unclear why this effect would no longer be at work in the high $T_{\mathrm{eq}}$ regime. \\
\indent That the theory matches best with observations at high $T_{\mathrm{eq}}$ is expected from \cite{Komacek:2015}, due largely to the very short radiative timescales at these high values of incident stellar flux. As a result, cloud formation is no longer an effective way to increase the phase curve amplitude in this regime. Additionally, although Lorentz forces are very strong in the high-temperature regime due to the steep dependence of thermal ionization on temperature \citep{Perna_2010_1,Rauscher_2013,Rogers:2014} the radiative timescale becomes so short that the observable atmosphere of the planet (at and above the photosphere) is close to radiative equilibrium. This forces large day-night temperature differences at the photosphere \citep{Komacek:2015}, as seen for our high-$T_{\mathrm{eq}}$ predictions in \Fig{fig:A_theory_individual}. 
\section{Double-grey numerical simulations}
\subsection{Numerical setup}
\label{sec:numsetup}
To understand the relevant dynamics governing the trend of increasing dayside-nightside temperature differences with increasing incident stellar flux found in \Sec{sec:comparisontoobs} in greater detail we turn to numerical (GCM) simulations. As in \cite{Komacek:2015}, we use the {\mitgcm} \citep{Adcroft:2004} to solve the hydrostatic primitive equations (Equations 1-5 in \citealp{Komacek:2015}). All aspects of the model dynamics, and the drag parameterization, Shapiro filter, and initial-temperature pressure profile are the same as in \cite{Komacek:2015}. However, instead of prescribing heating/cooling rates using a Newtonian cooling formalism as in the theory of \cite{Komacek:2015}, we couple a double-grey radiative transfer scheme, similar to that of \cite{Heng:2011a} and \cite{Rauscher_2012}, to the {\mitgcm} dynamical core. This scheme enables us to both test the qualitative predictions of \cite{Komacek:2015} using a more realistic model and produce simulated phase curves for comparison with observations. \\
\indent The plane-parallel, two-stream approximation of radiative transfer equations for the diffuse, azimuthally-averaged intensity $I$ are as follows (e.g. \citealp{Liou:2002}, Chapter 4.6), considering absorption only and omitting scattering,
\begin{equation}
\mu_1 \frac{d I^+}{d\tau} = I^+ - B(\tau),
\label{rteq1}
\end{equation}
\begin{equation}
-\mu_1 \frac{d I^-}{d\tau} = I^- - B(\tau),
\label{rteq2}
\end{equation}
where $I^{+}$ is the upward intensity and $I^{-}$ is the downward intensity of diffuse radiation in the infrared, $\tau$ is infrared optical depth, $\mu_1$ is the mean infrared zenith angle associated with a semi-isotropic hemisphere of radiation, and $B(\tau)$ is the Planck function at the local temperature $T(\tau)$. The double-grey assumption utilizes a separation between short wavelength (visible) insolation and long wavelength (thermal) emission. Equations (\ref{rteq1}) and (\ref{rteq2}) are solved to give the diffusive thermal flux $F_{\mathrm{th}}^{\pm} = \pi I_{\mathrm{th}}^{\pm}$, using boundary conditions of zero downward thermal flux at the top of atmosphere and a prescribed net upward thermal flux at the bottom of our domain. We use the hemi-isotropic (or hemispheric) closure for the thermal band, which assumes that the thermal radiation is isotropic in each hemisphere (upwelling and downwelling). This is a reasonable choice among the other possible closures used for the two-stream approximation, as it ensures energy conservation \citep{Pierrehumbert:2010}. \\
%As in \cite{Kylling:1995}, we take $\mu_{\mathrm{th}} = 0.5$. 
\indent We consider the stellar insolation as only a direct beam source without diffuse components, consistent with our assumption of no scattering. In this limit, the visible flux $F_\mathrm{v}^{\pm}$ is simply 
\begin{equation}
F_\mathrm{v}^- = \mu_\mathrm{v}F_0\exp(-\frac{\tau_\mathrm{v}}{\mu_\mathrm{v}}), \quad \quad F_\mathrm{v}^+ = 0,
\end{equation}
where $F_0$ is the visible flux incident on the top of the atmosphere at a given longitude and latitude, $\mu_\mathrm{v}$ the local zenith angle of the insolation, and $\tau_\mathrm{v}$ its optical depth at a given vertical location in the atmosphere. \\
\indent We utilize the reliable and efficient numerical package {\twostr} \citep{Kylling:1995} to solve the radiative transfer equations
% for a layered psuedo-spherical medium 
for a plane-parallel atmosphere in the two-stream approximation. {\twostr} is based on the general purpose multi-stream discrete ordinate algorithm {\disort} \citep{Stamnes:2027}, and incorporates all the advanced features of that well-tested and stable algorithm. One notable feature of {\twostr} is an efficient treatment of internal thermal sources that may be allowed to vary either slowly or rapidly with depth \citep{Kylling:1992}. The applicability of {\twostr} to atmospheric calculations has been tested thoroughly in \cite{Kylling:1995}. We repeated those tests with our version of the algorithm, and performed our own tests to ensure the correct implementation in the {\mitgcm} (see Appendix \ref{appendix} for an example).  \\
\indent To couple {\twostr} to the {\mitgcm}, the temperature-pressure structure of each vertical atmospheric column is fed from {\mitgcm} to calculate fluxes with {\twostr}. Then, we calculate the thermodynamic heating rate per unit mass, $q$, by taking the divergence in pressure coordinates of the net vertical flux $F$ as in \cite{Showmanetal_2009}
\begin{equation}
q = g\frac{\partial F}{\partial p},
\label{heatingrate}
\end{equation}
where $F = F_\mathrm{v}^+ - F_\mathrm{v}^- + F_{\mathrm{th}}^+ - F_{\mathrm{th}}^-$. This heating rate is then used for the next iteration of the dynamical equations. \\
\indent  In all of our numerical simulations varying incident stellar flux, drag strength, and rotation rate, we use visible and infrared opacities that do not change with these parameters. We do so to better understand the effects of incident stellar flux, drag strength, and rotation rate themselves have on the atmospheric circulation of hot Jupiters. This numerical exploration improves upon previous work exploring the how the atmospheric circulation varies with drag strength and day-night forcing amplitude using one-layer models \citep{Perez-Becker:2013fv} and 3D models with simplified Newtonian cooling \citep{Komacek:2015}. However, this model is not as complex as explorations in this parameter space with the full \sparc/\mitgcm \ model \citep{Showmanetal_2009,Kataria:2014,Kataria2016}, which use state-of-the-art radiative transfer with accurate opacities. As a result, this model is a middle rung in a ``modeling hierarchy'' exploring how the atmospheric circulation, especially the day-night temperature contrast, of hot Jupiters varies with incident stellar flux, drag strength, composition, and rotation rate.   \\
%\indent In all of our numerical simulations varying incident stellar flux, drag strength, and rotation rate, we use visible and infrared opacities that do not change with these parameters for simplicity. As a result, we will not directly compare the fractional day-night temperature differences to those predicted from our analytic theory that uses Newtonian cooling. At high values of incident stellar flux, the opacity should increase \citep{Freedman:2008,Freedman:2014}. Given that our opacity does not vary with incident flux, at these high stellar fluxes the infrared photosphere is at a greater pressure than it would if the opacity varied with temperature. As a result, our simulations have a lowered day-night temperature contrast at high $T_{\mathrm{eq}}$, as day-night temperature differences decrease with increasing pressure \citep{Komacek:2015} and our infrared photosphere is deeper than in a simulation with temperature-dependent opacities. However, we stress that our simulations are still useful for qualitative understanding of the atmospheric circulation, and can be used in \Sec{sec:consistentomega} to estimate how day-night temperature differences change with observed wavelength. \\ 
\indent We fix the visible absorption coefficient $\kappa_{\mathrm{v}} = 4 \times 10^{-4} \mathrm{m}^2 \ \mathrm{kg}^{-1}$ as in \cite{Rauscher_2012}. This is the same visible absorption coefficient used in the analytic solutions of \cite{Guillot:2010}, which is chosen to obtain a good match to the models of \cite{Fortney:2008} and \cite{Showman:2008}. Note that an increased value of the visible opacity (relevant to absorption by possible molecular TiO or VO) would cause a strong thermal inversion in the upper atmosphere, strongly affecting the temperature-pressure profile. We do not consider such an enhanced visible opacity in this work. We set the infrared absorption coefficient to be a power-law in pressure
\begin{equation}
\label{eq:kappapl}
\kappa_{\mathrm{th}} = c \left(\frac{p}{1 \ \mathrm{Pa}}\right)^b \mathrm{,}
\end{equation}
where $c = 2.28 \times 10^{-6} \mathrm{m}^2 \ \mathrm{kg}^{-1}$ and $b = 0.53$. This power-law opacity, which is commonly used to account for the effects of collision-induced or pressure-broadened absorption (e.g. \citealp{Arras:2006kl,Youdin_2010,Heng:2012b,Rauscher_2012}), gives the best-fit to the pressure-temperature profile from the analytic solutions of \cite{Parmentier:2014} and \cite{Parmentier:2014a} for HD 209458b using the full opacity-pressure-temperature relationship from \cite{Freedman:2008}.  \\
\indent We take the same planetary parameters for our numerical simulations as \cite{Komacek:2015}, see also \Sec{sec:theoryderiv}. We systematically vary the incident stellar flux in a sequence of models to determine its effect on the thermal structure and atmospheric circulation. The incident stellar flux can be described using an ``irradiation temperature,'' defined as $T_{\mathrm{irr}} = (F_{\star}/\sigma)^{1/4}$, where $F_{\star}$ is the incident stellar flux at the top of the atmosphere and $\sigma$ is the Stefan-Boltzmann constant. Thus, $T_{\mathrm{irr}}$ can be thought of as the equilibrium temperature of a blackbody at the substellar point. Equivalently, we can represent the stellar flux through the global-mean equilibrium temperature $T_{\mathrm{eq}}$. This is the temperature of a spherical blackbody planet where the heat is efficiently redistributed over all $4 \pi$ steradians, such that $T_{\mathrm{eq}} = T_{\mathrm{irr}}/4^{1/4}$. Note that in these simulations we assume zero planetary albedo such that $T_{\mathrm{eq}}$ is an effective control parameter. This is generally consistent with the observed low bond albedos for hot Jupiters \citep{Schwartz:2015}. 
%When varying the equilibrium temperature of the model, the input parameter for our numerical simulations is the irradiation temperature, $T_{\mathrm{irr}}$. The irradiation temperature is related to the equilibrium temperature as $T_{\mathrm{irr}} = 4^{1/4}T_{\mathrm{eq}}$. 
We set an internal temperature of $T_{\mathrm{int}} = 100 \ \mathrm{K}$, which is equivalent to an upward thermal flux at the bottom of the domain of $F^{+}_{\mathrm{th}} = \sigma T_{\mathrm{int}}^4$. This prescribed internal flux is small enough to not greatly affect our numerical solutions above the photosphere but large enough to aid in the equilibration of deep pressure levels of our GCM. \\
\indent In our numerical simulations, we use a horizontal resolution of C32 (approximately equal to a global resolution of $128 \times 64$ in longitude and latitude). There are 40 vertical levels, with the bottom 39 levels spaced evenly in log-pressure between $0.2 \ \mathrm{mbar} - 200 \ \mathrm{bars}$, and a top layer extending from $0.2 \ \mathrm{mbar}$ to space. We use a standard fourth-order Shapiro filter, which smooths grid-scale variations and enables the model to maintain numerical stability. We integrate our models until the circulation at photospheric pressures ($\lesssim 1 \ \mathrm{bar}$) reaches statistical equilibrium in thermal and kinetic energy. As a result, our simulations are each integrated to $2000 \ \mathrm{Earth } \ \mathrm{days}$ of model time, and our results are time-averaged over the last $100 \ \mathrm{days}$ of model time. 
%\indent In the remainder of the section, we analyze our numerical results. In \Sec{sec:trendsnumerical}, we examine a suite of simulations varying equilibrium temperature and drag strength, and in \Sec{sec:consistentomega} we focus on simulations with consistently varying equilibrium temperature and rotation rate.
\subsection{Trends in atmospheric circulation with varying stellar irradiation and drag strengths}
\label{sec:trendsnumerical}
\begin{figure*}
	\centering
	\includegraphics[width=0.99\textwidth]{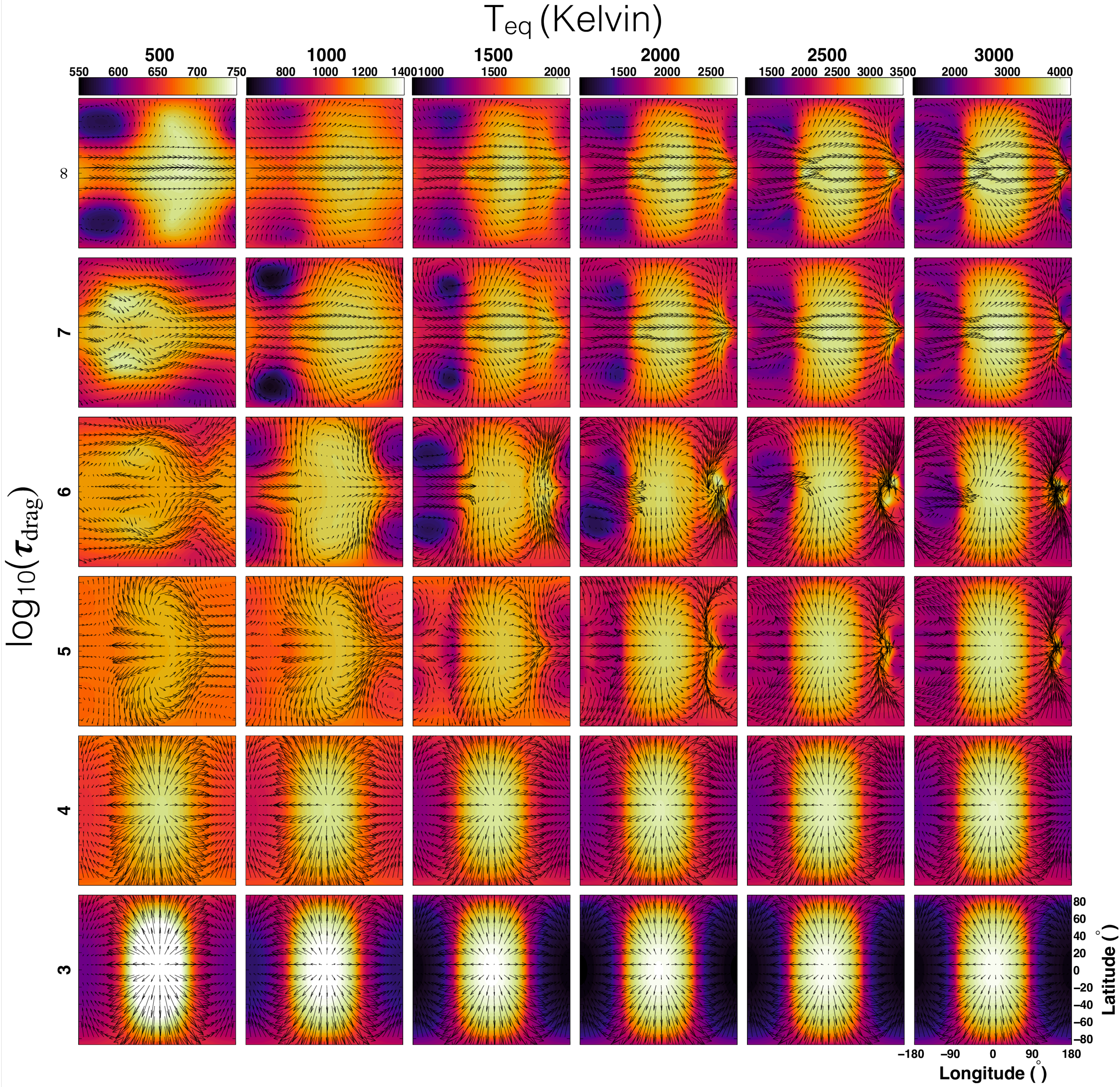}
	\caption{Maps of temperature (colors, in K) and wind (vectors) at $80 \ \mathrm{mbar}$ pressure for 36 separate GCM simulations varying the incident stellar flux (with values of $1.418 \times 10^4, 2.268 \times 10^5, 1.148 \times 10^6, 3.628 \times 10^6, 8.859 \times 10^6, 1.837 \times 10^7 \ \mathrm{W}\mathrm{m}^{-2}$), corresponding to varying global-mean equilibrium temperature from $T_{\mathrm{eq}} = 500 - 3000 \ \mathrm{K}$. Additionally, we vary the drag timescale $\tau_{\mathrm{drag}}$ from $10^3 \ \mathrm{s} - \infty$. Each column shares a color scale, with brighter (darker) colors corresponding to hotter (colder) temperatures. Each plot has independent horizontal wind vectors, meant to show the geometry of the flow. These simulations include band-grey radiative transfer, using a constant visible opacity, and an infrared opacity set to be a function of pressure alone. As $T_{\mathrm{eq}}$ is increased, dayside-nightside temperature differences increase. Additionally, if $\tau_{\mathrm{drag}}$ is short ($10^3 - 10^5 \ \mathrm{s}$), equatorial superrotation is cut off and dayside-nightside temperature differences become large.}
	\label{fig:tempgrid}
\end{figure*}
Our main suite of numerical simulations examine how the atmospheric circulation changes with varying incident stellar flux and drag strengths (parameterized by a drag timescale $\tau_{\mathrm{drag}}$). This is comparable to our non-linear suite of simulations varying radiative and drag timescales in \cite{Komacek:2015}, except here we solve the actual radiative transfer equations (albeit in a double-grey two-stream approximation) rather than using Newtonian cooling. We use the same drag scheme as in \cite{Komacek:2015}, which consists of two parts. First, we include a weak frictional drag at the bottom of the model, extending from $200 \ \mathrm{bars}$ to $10 \ \mathrm{bars}$. The drag coefficient for this component is largest (while still being relatively weak) at the bottom, and decreases with decreasing pressure, becoming zero (i.e. with no applied drag) at and above $10 \ \mathrm{bars}$. Second, we optionally add a spatially constant drag term with a drag timescale of $\tau_{\mathrm{drag}}$. The total drag coefficient at each pressure level is taken to be the greater of the individual drag coefficients from these two components (see \citealp{Komacek:2015}, Equation 12). \\
% with weak basal drag extending from $10 \ \mathrm{bars}$ to the bottom of the domain. If atmospheric drag is considered, it has the same strength throughout the atmosphere. 
\indent As in \cite{Komacek:2015}, we examine simulations with varying $\tau_{\mathrm{drag}}$ from $10^3 - \infty \ \mathrm{s}$. Simulations with $\tau_{\mathrm{drag}} = \infty$ have no large-scale drag at pressures less than $10 \ \mathrm{bars}$, but they (and all other simulations presented in this paper) still have the basal drag scheme at the bottom of the domain. We also vary the incident stellar flux at the substellar point (with values of $1.418 \times 10^4, 2.268 \times 10^5, 1.148 \times 10^6, 3.628 \times 10^6, 8.859 \times 10^6, 1.837 \times 10^7 \ \mathrm{W}\mathrm{m}^{-2} $), corresponding to varying global-mean equilibrium temperatures $T_{\mathrm{eq}}$ from $500 - 3000 \ \mathrm{K}$ in steps of $500 \ \mathrm{K}$. This results in a grid of $36$ simulations, with resulting temperature and wind maps at $80 \ \mathrm{mbar}$ shown in \Fig{fig:tempgrid}. \\
\indent The results of this suite of simulations agree well with both theoretical expectations and the simulations using Newtonian cooling in \cite{Komacek:2015}, compare \Fig{fig:tempgrid} here to their Figure 4. Dayside-nightside temperature differences are small at low $T_{\mathrm{eq}}$ and become larger with increasing $T_{\mathrm{eq}}$, as expected from the theory in \Sec{sec:comparisontoobs}. Additionally, when drag is strong, with a drag timescale $\tau_{\mathrm{drag}} \lesssim 10^4 \ \mathrm{s}$, characteristic dayside-nightside temperature differences are larger than in simulations with weaker drag. This is expected from the theory in \cite{Komacek:2015}, as drag should increase the day-night temperature contrast if the condition $\tau_{\mathrm{drag}} \ll \Omega^{-1}$ is satisfied. Given that in these simulations we use a rotation period of $3.5 \ \mathrm{days}$ and hence $\Omega^{-1} = 4.8 \times 10^4 \ \mathrm{s}$, we expect that a transition between low and high day-night temperature differences occurs between $\tau_{\mathrm{drag}} = 10^5 \ \mathrm{s}$ and $\tau_{\mathrm{drag}} = 10^4 \ \mathrm{s}$, and this prediction agrees with our current grid of simulations. \\
\indent As in \cite{Komacek:2015}, our simulations show a change between long and short $\tau_{\mathrm{drag}}$ in the general character of the atmospheric circulation near the infrared photosphere. At long $\tau_{\mathrm{drag}} > 10^5 \ \mathrm{s}$, there is a dominant eastward jet at the equator. At short $\tau_{\mathrm{drag}}$, however, the flow is predominantly from day to night, rather than east-west. This transition may be observationally detectable \citep{parmentier_2013,showman_2013_doppler}, and indeed the Doppler analysis of HARPS spectra by \cite{Louden:2015} found that the atmospheric flow of HD 189733b is likely dominated by an east-west jet, rather than day-night flow. \\
\indent A subset of our simulations with $\tau_{\mathrm{drag}} = 10^6 \ \mathrm{s}$ and high $T_{\mathrm{eq}}$ do not show mirror symmetry of the wind pattern about the equator. These simulations are time-variable, with the eastward jet notably fluctuating in latitude at the location west of the substellar point where a hydraulic jump has been seen in previous simulations (e.g. \citealp{Showmanetal_2009}). Recent work by \cite{Fromang:2016} has shown that shear instabilities can cause hot Jupiter atmospheres to have (potentially observable) time variability, as predicted by \cite{showman_2002,Goodman:2009,Li:2010}. The GCMs of \cite{Showmanetal_2009} predicted on the order of $1 \%$ variability in the secondary eclipse depths of HD 189733b, in agreement with the upper limit on variability found from the observations of \cite{Agol:2010}. We plan to quantify the observability of time-variability in future work, as it has the potential to affect observations of exoplanet atmospheres with \textit{JWST}. 
%stuff about winds, doppler paper, time variability hint \\
%Describe Figure 4 in detail, and how it matches up with thoretical expectations from Figure 2. Explain how wave pattern changes with changing drag and radiative timescales (showman and polvani, doppler paper) , re-iterate wave adjustment point. 
\subsection{Simulated phase curves}
\label{sec:phasecurvecalc}
\begin{figure*}
	\centering
	\includegraphics[width=1\textwidth]{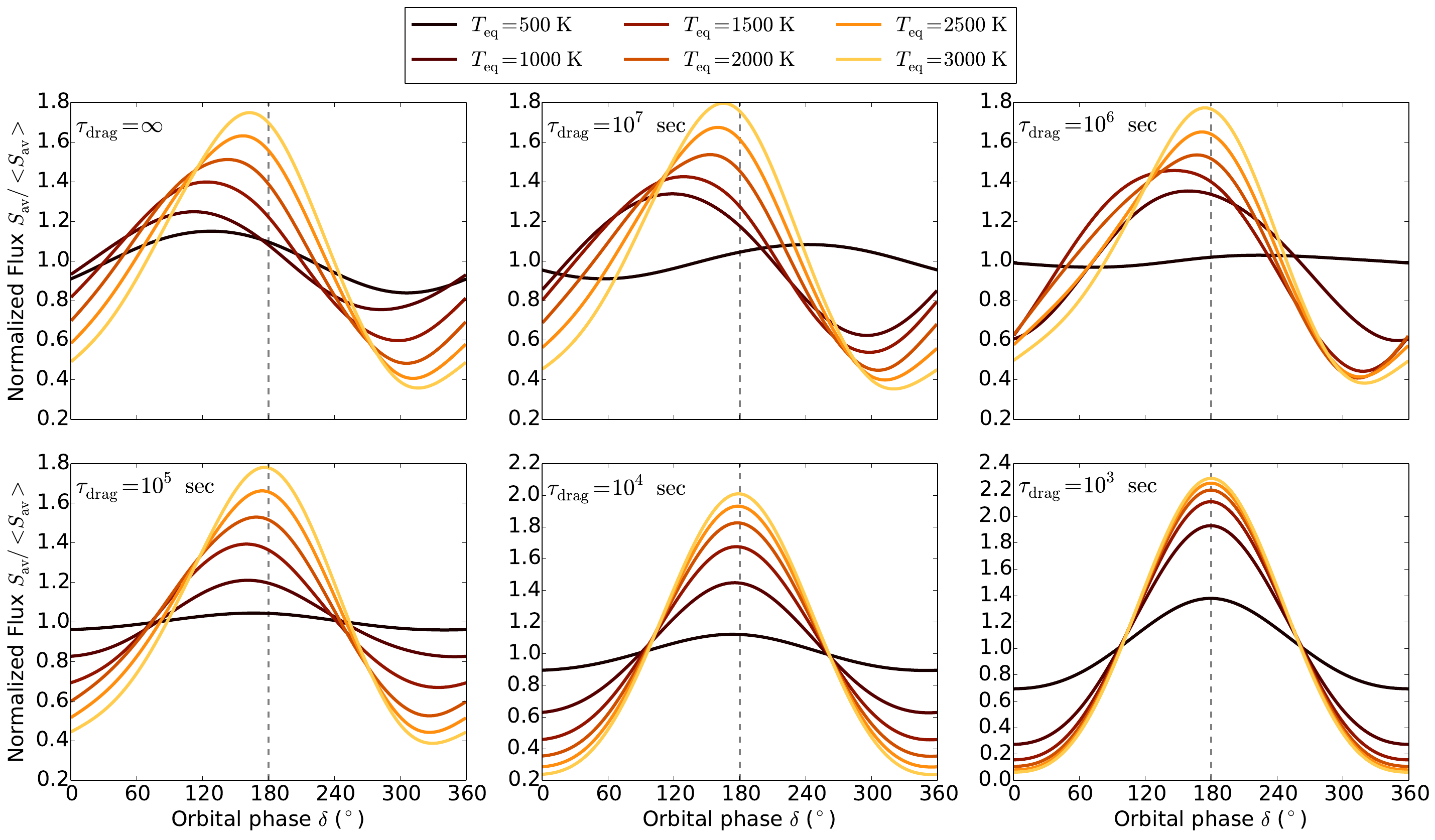}
	\caption{Broadband infrared phase curves for each of the GCMs with double-grey radiative transfer shown in \Fig{fig:tempgrid}. Primary transit occurs at phase $0^{\circ}$ and secondary eclipse (shown by the dashed line) occurs at $180^{\circ}$. The outgoing flux is normalized by the orbit-averaged flux of each phase curve. Each sub-plot shows how the phase curve varies with varying $T_{\mathrm{eq}}$ for a given value of $\tau_{\mathrm{drag}}$. Note that the last two panels (for $\tau_{\mathrm{drag}} = 10^4 \ \mathrm{and} \ 10^3 \ \mathrm{s}$) use a different y-axis scale due to the significantly larger phase curve amplitude for these short drag time constants. We calculate the flux for each phase curve by post-processing each model with \twostr \ to find the flux at each grid point and then averaging over the Earth-facing hemisphere using \Eq{eq:flux} to find the mean flux the planet is radiating toward Earth as a function of orbital phase. Generally, the phase curve amplitude for each light curve increases with increasing $T_{\mathrm{eq}}$ and decreasing $\tau_{\mathrm{drag}}$, while the phase offset decreases with increasing $T_{\mathrm{eq}}$ and decreasing $\tau_{\mathrm{drag}}$.}
	\label{fig:lightcurves}
\end{figure*}
We post-process the resulting temperature-pressure profiles from our GCMs using \twostr \ in order to calculate light curves as a function of orbital phase. Specifically, we calculate the mean flux that the Earth-facing hemisphere radiates toward Earth, which is a function of orbital phase. Here we write this hemispheric-mean flux radiated towards Earth as $S_{\mathrm{av}}(\delta)$, where $\delta$ is the orbital phase (taken to be equivalent to the sub-Earth longitude). We assume a circular orbit and zero obliquity, causing the orbital phase to vary linearly with time through the orbit. Here we use the same definition for orbital phase as in \cite{Zhang:2016}, such that $\delta$ is equal to $0^{\circ}$ at primary eclipse and $180^{\circ}$ at secondary eclipse. Similarly, $\delta = 90^{\circ}$ and $\delta = 270^{\circ}$ correspond to times in the orbit where the eastern and western terminators of the planet lie at the sub-Earth longitude, respectively. Additionally, note that as in \cite{Zhang:2016} we define the planetary longitude (here written as $\lambda$) such that $\lambda = 0^{\circ}$ at the substellar point and $\lambda = 180^{\circ}$ at the anti-stellar point.   \\
\indent To calculate the local outgoing (top-of-atmosphere) flux in the direction of Earth at each grid point and at each orbital phase (here written $S(\lambda,\phi,\delta)$) we use the local temperature-pressure profile as a function of latitude (here written as $\phi$) and longitude from the same GCM results as shown in \Fig{fig:tempgrid} that have been time-averaged over the last 100 days of integration. We then use \twostr \ to calculate the outgoing flux using the particular value of $\mu$ relevant for the latitude and longitude of each grid point, along with taking into account the orbital phase. Here $\mu(\lambda,\phi,\delta)$ (similar to the visibility in \citealp{Cowan2009}) is the cosine of the angle between the local normal to each grid point and the line of sight toward Earth, taking into account both the angle between the longitude of each grid point and the sub-Earth longitude and the angle between the latitude of each grid point and the equator. As a result, this post-processing approach correctly accounts for the directionality of the emitted radiation from the planet.  \\
\indent We then integrate up the outgoing flux radiated towards Earth at each grid point to calculate a hemispheric-mean flux radiated towards Earth. This approach is the same as that used in previous work (e.g. \citealp{Fortney:2006a,Cowan2009,Showman:2008,Showmanetal_2009,Lewis:2010,Kataria:2014,Showman:2014}), computing the integral
%Using the outgoing (top of atmosphere) infrared flux computed at each longitude and latitude from our GCM ($F(\lambda,\phi)$ where $\lambda$ is longitude and $\phi$ latitude) and combining the contribution from each grid point to the line of sight observed flux, we can compute infrared light curves. To do so, we compute the following integral to find the line of sight flux $F_{\mathrm{LOS}}$ at each at each longitude:
\begin{equation}
\label{eq:flux}
S_{\mathrm{av}}(\delta) = \frac{1}{\pi a^2} \int^{+\pi/2}_{-\pi/2} \int^{3\pi/2-\delta}_{\pi/2-\delta} S(\lambda,\phi,\delta) \mu(\lambda,\phi,\delta) dA \mathrm{,}
\end{equation}
where $dA = a^2\mathrm{cos}(\phi)d\phi d\lambda$ with $d\phi $ and $d\lambda$ the subtended latitudinal and longitudinal angles of each grid point, respectively. \\
\indent The resulting phase curves calculated from each of our GCMs using \Eq{eq:flux} are shown in \Fig{fig:lightcurves}. Here we have normalized the flux for each phase curve by the average flux radiated toward Earth over an orbit. There are two key parameters we can pull out of these simulated phase curves: the phase curve amplitude and the phase offset of the peak emitted flux from secondary eclipse. In general, as expected from our theory, the phase curve amplitude increases with increasing $T_{\mathrm{eq}}$ and decreasing $\tau_{\mathrm{drag}}$. The phase offset shows the opposite trend, generally decreasing with increasing $T_{\mathrm{eq}}$ and decreasing $\tau_{\mathrm{drag}}$. Next, we will directly compare our simulated phase curve amplitudes to observations in \Sec{sec:bandresults}, and later will explore trends in phase offsets further in \Sec{sec:phaseoffsets}. 
\subsection{Numerical phase curve amplitudes}
%\subsection{Expected trends in dayside-nightside temperature difference with observed wavelength}
\label{sec:bandresults}
\begin{figure}
	\centering
	\includegraphics[width=0.5\textwidth]{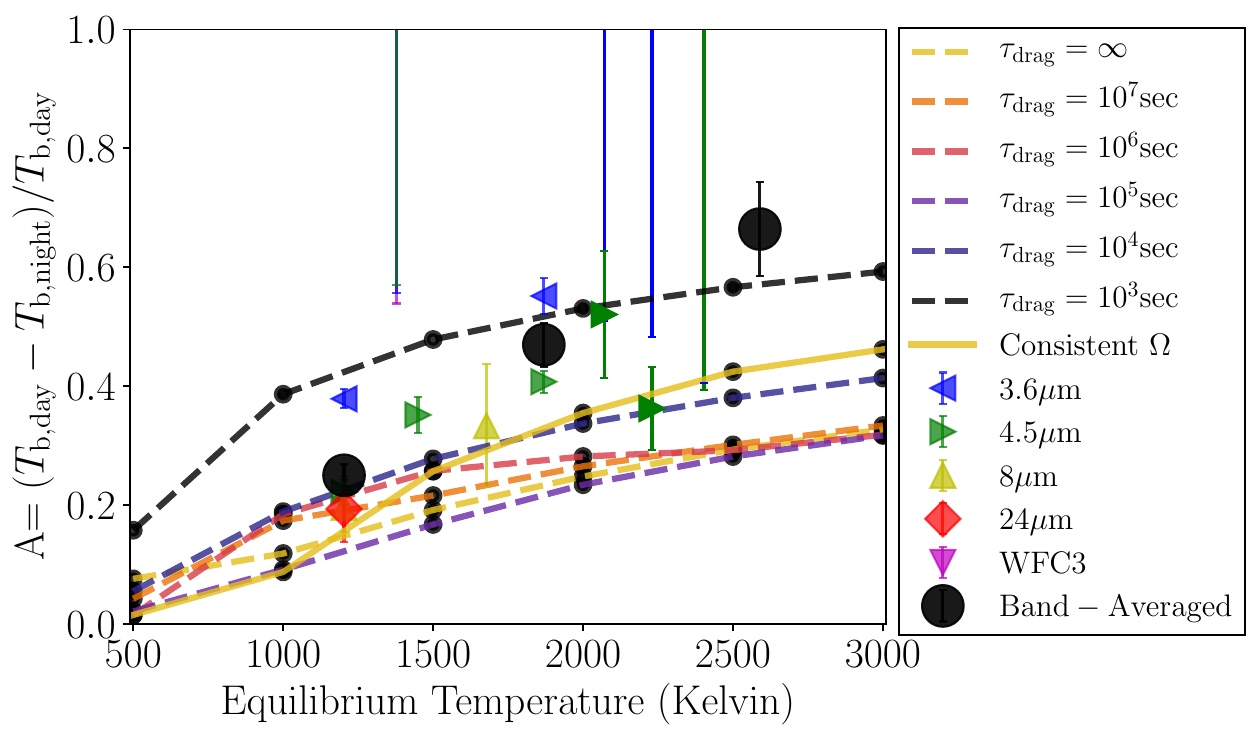}
	\caption{Comparison between double-grey numerical GCM calculations (lines, dots) and observations (solid points) of fractional dayside-nightside brightness temperature differences (or phase curve amplitudes) for each individual low-eccentricity transiting planet with full-phase infrared observations. Small black dots show the numerical fractional dayside-nightside brightness temperature difference itself, while dashed lines connect these points for our synchronously rotating simulations with a constant rotation period of $3.5 \ \mathrm{days}$. Observations are the same as those shown in Figure \ref{fig:Aobs}. The solid line and dots show results from our simulations with consistent rotation rate and stellar irradiation from \Sec{sec:consistentomega}. A similar trend of increasing phase curve amplitude with increasing incident stellar flux as seen in the observations and theory from \Sec{sec:comparisontoobs} is seen in the numerical simulations here. However, studying the atmospheric dynamics using GCMs shows subtle changes in the dayside-nightside temperature contrast with changing $\tau_{\mathrm{drag}}$ not seen in the analytic predictions. Notably, the $\tau_{\mathrm{drag}} = 10^5 \ \mathrm{s}$ case shows the lowest phase curve amplitudes, not the $\tau_{\mathrm{drag}} = \infty$ case. The numerical day-night temperature differences are smaller than the theoretical predictions at high $T_{\mathrm{eq}}$ because here we keep the opacities fixed with temperature, when they should in fact increase.}
	\label{fig:A_numerical}
\end{figure}
\indent Picking off the maximum and minimum of the mean flux radiated toward Earth and calculating their effective temperatures $T_{\mathrm{eff}} = (S_{\mathrm{av}}/\sigma)^{1/4}$, we can calculate the phase curve amplitude $A$ from our numerical simulations.
% We can also compute the offset of the point of maximum flux from the substellar point, which will be discussed in \Sec{sec:phaseoffsets}. 
Note that for our double-grey simulations, which only have one infrared band, the effective temperature is equivalent to the brightness temperature. These numerical predictions of the phase curve amplitude for varying equilibrium temperature and drag strengths are shown (along with the observations from \Fig{fig:Aobs}) in \Fig{fig:A_numerical}. \\
\indent As expected from the theoretical predictions in \Sec{sec:comparisontoobs} and as seen in observations, the numerical phase curve amplitudes increase with increasing incident stellar flux. Additionally, the presence of strong drag can greatly increase the phase curve amplitude. As expected from the theoretical predictions, the day-night temperature difference is significantly larger for the models with $\tau_{\mathrm{drag}} < 10^5 \ \mathrm{s}$ than those with $\tau_{\mathrm{drag}} \ge 10^5 \ \mathrm{s}$. However, the exact dependence of the phase curve amplitude on $\tau_{\mathrm{drag}}$ for the cases with $\tau_{\mathrm{drag}} \ge 10^5 \ \mathrm{s}$ is subtle, as the $\tau_{\mathrm{drag}} = 10^5 \ \mathrm{s}$ case actually shows smaller phase curve amplitudes than the $\tau_{\mathrm{drag}} = \infty$ model. This is because the $\tau_{\mathrm{drag}} = 10^5 \ \mathrm{s}$ is a transition case between an equatorial jet and dayside-nightside flow, and lacks the cold mid-latitude vortices seen on the nightside of the cases with $\tau_{\mathrm{drag}} \ge 10^6 \ \mathrm{s}$ (see \Fig{fig:tempgrid}). Lastly, note that the trend of increasing $A$ with increasing $T_{\mathrm{eq}}$ is weaker for our numerical simulations than for our theoretical predictions (compare with \Fig{fig:A_theory}). This is likely due to the lack of an opacity temperature-dependence in our numerical simulations, which if included would shift the photosphere to higher pressures at higher equilibrium temperatures and increase the simulated phase curve amplitudes. 
\subsection{Numerical simulations with consistent rotation rate and stellar irradiation}
\label{sec:consistentomega}
\begin{figure*}
	\centering
	\includegraphics[width=1\textwidth]{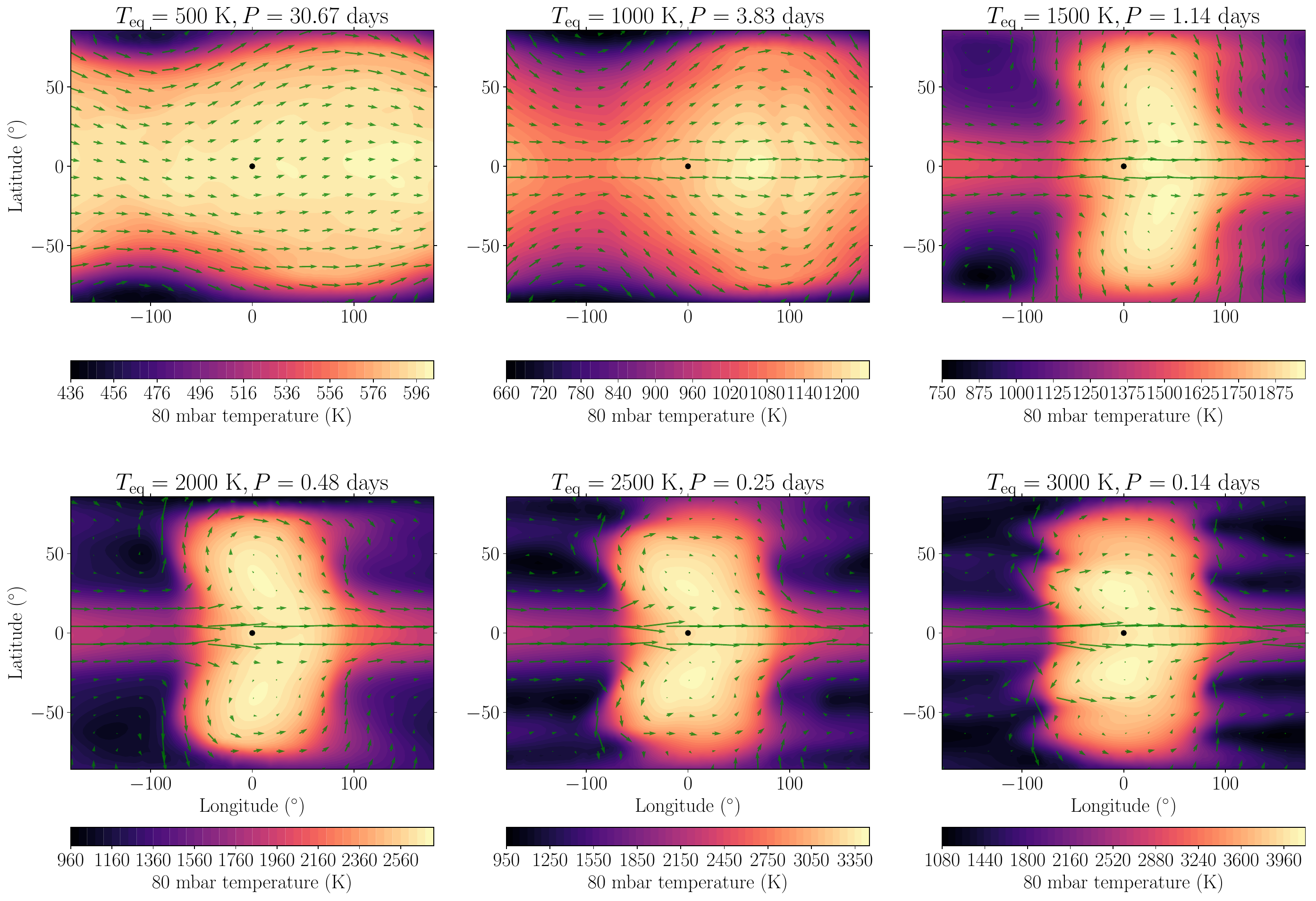}
	\caption{Maps of temperature (colors, in K) and wind (vectors) at $80 \ \mathrm{mbar}$ pressure for 6 separate GCM simulations varying the equilibrium temperature from $500 - 3000 \ \mathrm{K}$ while keeping the rotation period equal to the orbital period the planet would have if it was orbiting a star with the effective temperature and radius of HD 189733. As a result, the rotational period for each simulation varies from $30.67$ to $0.14 \ \mathrm{days}$, decreasing with increasing $T_{\mathrm{eq}}$. Each plot has an individual color scale and overplotted horizontal wind vectors.} 
	%Equatorial superrotation occurs in all simulations with $T_{\mathrm{eq}} \ge 1000 \ \mathrm{K}$. As seen in \Fig{fig:A_numerical}, the fractional dayside-nightside temperature differences in these models do not have a steeper relationship with $T_{\mathrm{eq}}$ than our simulations with a constant rotation rate. Instead, they have very similar phase curve amplitudes at high $T_{\mathrm{eq}} = 2500-3000 \ \mathrm{K}$, and slightly smaller amplitudes at $T_{\mathrm{eq}} \le 2000 \ \mathrm{K}$. This is likely because the nightside vortices in these simulations have a smaller contribution toward lowering the flux emitted from the nightside and hence increasing the phase curve amplitude. Specifically, the nightside vortices are at higher latitudes here than those in the grid shown in \Fig{fig:tempgrid}, and as a result these regions have a lower emitted flux toward Earth in this grid of simulations with a consistent rotation rate than in the larger grid with a constant rotation rate.}
	%Fractional dayside-nightside temperature differences are low for the cases with $T_{\mathrm{eq}} = 500, 1000 \ \mathrm{K}$, but large for all simulations with $T_{\mathrm{eq}} \ge 1500 \ \mathrm{K}$.}
	\label{fig:tempgrid_omega}
\end{figure*}
\indent We ran a secondary suite of GCMs varying $T_{\mathrm{eq}}$ from $500 - 3000 \ \mathrm{K}$ and fixing $\tau_{\mathrm{drag}} = \infty$, while keeping the rotation period at the value it would have if it was equal to the orbital period of a planet orbiting a star with the same bolometric luminosity as HD 189733. We do so in order to analyze the effects of rotation rate on our results from \Sec{sec:trendsnumerical}, as in our main grid of simulations we kept the rotation rate fixed. Note that this set of models is similar to the double-grey GCMs performed in \cite{perna_2012}, but here we span a slightly larger range of incident stellar flux. The relationship between the orbital period $P_{\mathrm{orb}}$ and $T_{\mathrm{eq}}$ for a blackbody with zero albedo is
\begin{equation}
P_{\mathrm{orb}} = \frac{\pi^{1/4}}{4} \left(\frac{L_{\star}}{\sigma}\right)^{3/4} \left(GM_{\star}\right)^{-1/2} T_{\mathrm{eq}}^{-3} \mathrm{,}
\end{equation}  
where $L_{\star}$ and $M_{\mathrm{\star}}$ are the stellar luminosity and mass, respectively, and $G$ is the gravitational constant. We calculate $L_{\star} = 4 \pi R^2_{\mathrm{\star}} \sigma T^4_{\star}$, where $R_{\mathrm{\star}}$ and $T_{\star}$ are the stellar radius and effective temperature, respectively, and we have assumed that the stellar spectrum can be approximated as a blackbody. Here we take values of stellar radius, mass, and effective temperature relevant to HD 189733: $R_{\mathrm{\star}} = 0.805 R_{\varodot}$, $M_{\mathrm{\star}} \approx 0.8 M_{\varodot}$, and $T_{\star} = 4875 \ \mathrm{K}$. This results in orbital and rotation periods decreasing from $30.67 - 0.14$ days as $T_{\mathrm{eq}}$ increases from $500 - 3000 \ \mathrm{K}$. Note that the short end of these orbital periods corresponds to a distance from the star of $\sim 2$ stellar radii. As a result, our simulations with very large $T_{\mathrm{eq}}$ are included to understand the expected behavior of the circulation in the high-$\Omega$ limit. However, note that hot Jupiters with such short orbital periods would likely be destroyed due to tidal interactions with their host star. Additionally, planets with long orbital periods are not expected to be tidally locked, and hence this suite of simulations was run solely to understand the change in dynamics with varying $T_{\mathrm{eq}}$ and $\Omega$, not to compare with observations. \\
\indent The results of the suite of simulations consistently varying $T_{\mathrm{eq}}$ and $\Omega$ are shown in \Fig{fig:tempgrid_omega}. First, note that the atmospheric circulation for $T_{\mathrm{eq}} \ge 1000 \ \mathrm{K}$ is in general similar to the models in \Sec{sec:trendsnumerical}, with an eastward equatorial jet dominating the flow pattern. For comparison with the GCMs with fixed rotation rate, we show the dayside-nightside temperature contrast from this grid of simulations along with the results from the larger grid in \Fig{fig:A_numerical}. As in the simulations with consistent rotation rate from \cite{perna_2012}, the fractional dayside-nightside temperature differences increase with increasing $T_{\mathrm{eq}}$. This is the same trend expected from our theory and found from our grid of simulations with a constant rotation rate shown in \Fig{fig:tempgrid}. \\
%This is expected from theory and the grid of simulations in \Fig{fig:tempgrid}. 
%\indent The fractional dayside-nightside temperature differences at high $T_{\mathrm{eq}} \ge 2500 \ \mathrm{K}$ match well with those from the simulations with fixed rotation rate. However, the fractional dayside-nightside temperature differences at intermediate $1000 \ \mathrm{K} \le T_{\mathrm{eq}} \le 2000 \ \mathrm{K}$ are significantly lower than those from the models with a constant rotation rate. This is likely because of the decreased effect of the mid-latitude nightside vortices on increasing the day-night temperature contrast in our runs with consistent rotation rate. This is because the vortices occur at higher latitudes in these simulations with a consistently varying rotation rate than in the simulations with a constant rotation rate.  \\
%However, the dayside-nightside temperature differences at high-$T_{\mathrm{eq}}$ are slightly larger in \Fig{fig:tempgrid_omega} than in \Fig{fig:tempgrid}, as faster rotation rates increase dayside-nightside temperature differences (\Sec{sec:indivpred}). 
\indent Note that in the consistent-rotation simulations with high $T_{\mathrm{eq}} \ge 2000 \ \mathrm{K}$, the temperature and wind patterns are very similar. At such high $T_{\mathrm{eq}}$, the maximum zonal-mean zonal wind speed begins to saturate at $\sim 6.5 \ \mathrm{km} \ \mathrm{s}^{-1}$. Additionally, the simulations with short rotation periods have a larger equator-to-pole temperature contrast than those in \Fig{fig:tempgrid}. This is because, as discussed by \cite{Showman:2014}, more rapidly rotating flows can support larger horizontal temperature differences. This is because the increasing Coriolis force with increasing rotation rate can support larger pressure gradients, which are related to the latitudinal temperature difference (see Equation 8 of \citealp{Showman:2014}). 
%This is expected, as the Hadley circulation transporting heat from equator to pole decreases in latitudinal width and efficacy with increasing $\Omega$ \citep{Held:1980}. 
%This is because the increasing rotation rate decreases the efficacy of wave action to drive the equatorial jet \citep{Showman_Polvani_2011}, offsetting the effects of increasing day-night forcing amplitude. 
%Discuss how day-night temperature differences and wind speeds saturate at high omega (even at unphysical rotation periods). Discuss qualitative differences between runs with consistent rotation rate. geopotential high anticyclonic
\subsection{Infrared phase offsets}
\label{sec:phaseoffsets}
\subsubsection{Numerical results}
\label{sec:numericaloffset}
\begin{figure}
	\centering
	\includegraphics[width=0.5\textwidth]{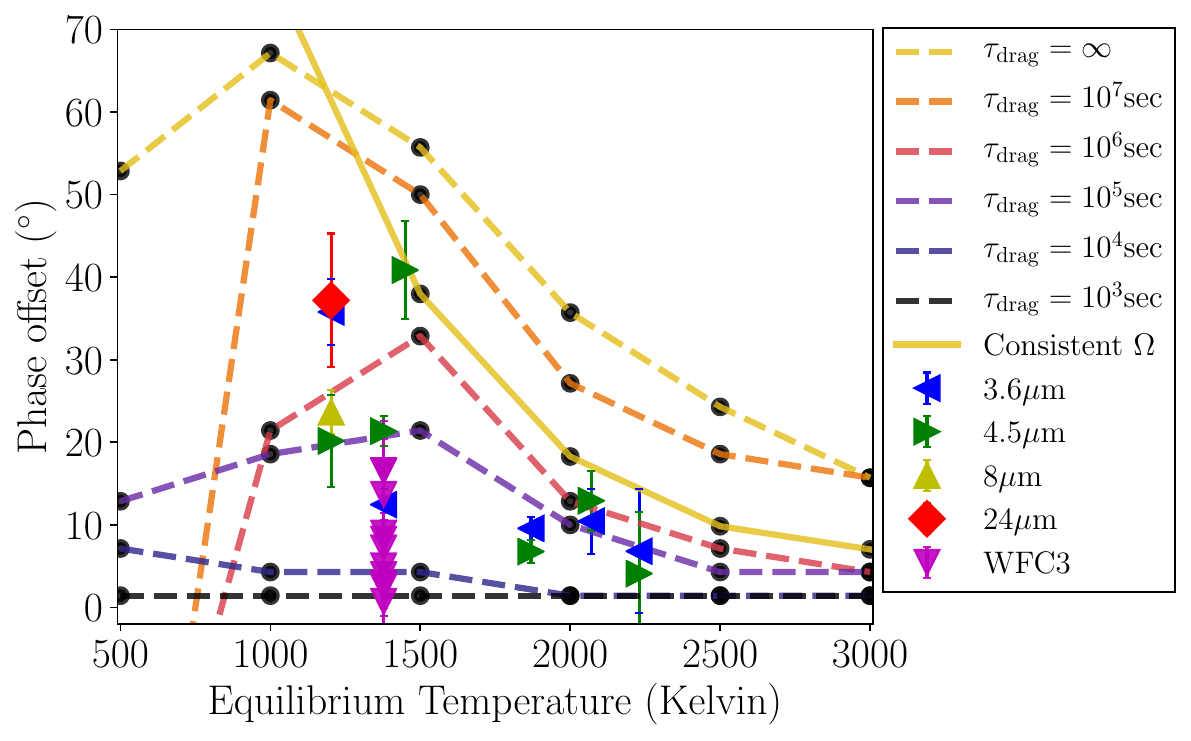}
	\caption{Observed infrared phase offsets in given infrared wavebands (large solid points) and phase offsets from double-grey numerical (GCM) simulations (dashed lines, dots) varying $\tau_{\mathrm{drag}}$ plotted against global equilibrium temperature. As in \Fig{fig:A_numerical}, we also show the results from our grid of GCMs with consistent rotation rate (solid line, dots). A positive phase offset corresponds to an eastward shift in the infrared ``hot spot'' of the planet. Black dots show the phase offsets from the simulations themselves, while dashed lines connect these results. Observations are taken from the same references as in \Fig{fig:Aobs}. We plot all of the individual WFC3 bands from the observations of \cite{Stevenson:2014} in order to show the variations in phase offset within the WFC3 wavelength range.  Note that the phase offsets for the GCMs with consistent rotation rate for $T_{\mathrm{eq}} = 500$ and $1000 \ \mathrm{K}$ lie above the y-axis cutoff and are $80.16^{\circ}$ and $77.34^{\circ}$, respectively. The simulations of \cite{perna_2012} with a consistent rotation rate also found large phase offsets at low values of incident stellar flux.
	% For clarity, the observations are colored by the global equilibrium temperature of the planet. Most observations are from the same references as in Figure \ref{fig:Aobs}, for details see the review of \cite{Crossfield:2015}. We have updated their data to include the observations from \cite{Wong:2015a} of WASP-19b and HAT-P-7b. 
Both the observations and GCM results show the same general trend of decreasing phase offset with increasing equilibrium temperature. Using the phase offsets from the GCM, some degree of atmospheric drag (ranging from drag timescales of $10^4 - 10^7 \ \mathrm{s}$) is needed to explain each of the observed infrared phase offsets. 
%However, given the strong drag needed to explain the small phase offsets observed for the two coolest planets in our sample (HD 189733b and WASP-43b) it is unlikely that that drag alone can explain these observations.
}
	\label{fig:phaseshift_numerical}
\end{figure}
\indent Post-processing our numerical results from \Sec{sec:trendsnumerical} using the method in \Sec{sec:phasecurvecalc}, we obtain the offset of the peak of the infrared phase curve along with the amplitude discussed previously. Here, we calculate the offset from secondary eclipse for each broadband infrared phase curve from \Fig{fig:lightcurves} in order to compare with observational data. \Fig{fig:phaseshift_numerical} shows a comparison between our numerically calculated infrared phase curve offsets as a function of $T_{\mathrm{eq}}$ for varying $\tau_{\mathrm{drag}}$, along with the observed offsets of the peak of the infrared phase curve for the same low-eccentricity transiting hot Jupiters as in \Fig{fig:Aobs}. The general trend resulting from our numerical simulations is decreasing eastward phase offsets with increasing $T_{\mathrm{eq}}$, as seen from the observations and expected from the theory of \cite{Cowan:2011} and also from \cite{Zhang:2016}. We will compare directly our numerical results with the analytic theory of \cite{Zhang:2016} in \Sec{sec:compareoffset}.   \\
\indent Our numerical simulations capture the general trend of observed decreasing infrared phase offsets with increasing incident stellar flux and can explain the entire set of observed infrared phase offsets with varying values of $\tau_{\mathrm{drag}}$. However, there is considerable scatter in the observed infrared phase offsets between different wavelength bands, so in the context of our numerical simulations a wide range ($10^4 - 10^7 \ \mathrm{s}$) of drag timescales must be invoked to explain the observed phase shifts. Considering Lorentz forces as the dominant drag mechanism, this would only be physical if hot Jupiters had a wide range of magnetic field strengths or atmospheric compositions and hence ionization fractions. The former is unlikely if the magnetic field is driven by an internal dynamo \citep{Christensen:2009,Christensen:2010}. The latter possibility may have an effect, but note that sodium (the element with the lowest ionization potential) is found in a variety of hot Jupiters regardless of equilibrium temperature \citep{Sing:2015a}. Additionally, note that the Rayleigh drag formulation that we use is at best a very rough approximation to the true effects of Lorentz forces \citep{Perna_2010_1,Rauscher_2013,Heng:2014,Rogers:2014}. We will discuss this further in \Sec{sec:discussion}. \\ 
\indent More generally, it is theoretically expected that a competition between the speed of the equatorial jet and the radiative cooling of the circulation determines the phase offset \citep{Cowan:2011,Zhang:2016}. Given that drag on the winds acts to slow the equatorial jet (both directly through decelerating the winds and indirectly by affecting the wave action that drives the jets), invoking a variety of $\tau_{\mathrm{drag}}$ is sufficient to explain observations in the context of our numerical simulations. However, as discussed above, this is likely unphysical for the case of magnetic drag. 
%Additionally, if drag of the jet is due to shear instabilities
%Additionally, if drag of the jet is due to shear instabilities (e.g. as seen in the simulations of \citealp{Fromang:2016}), one would expect that the effective drag strength would increase with increasing jet speeds and hence incident stellar flux. 
In general, it is unclear exactly how the effective drag strength should vary with planetary parameters, largely because the most important drag mechanism in hot Jupiter atmospheres has yet to be identified. Future observations could help to determine whether the observed scatter in infrared phase offsets is real, or if there is instead a clear trend in phase offset with incident stellar flux. 
\subsubsection{Comparison with analytic theory}
\label{sec:compareoffset}
\begin{figure}
	\centering
	\includegraphics[width=0.5\textwidth]{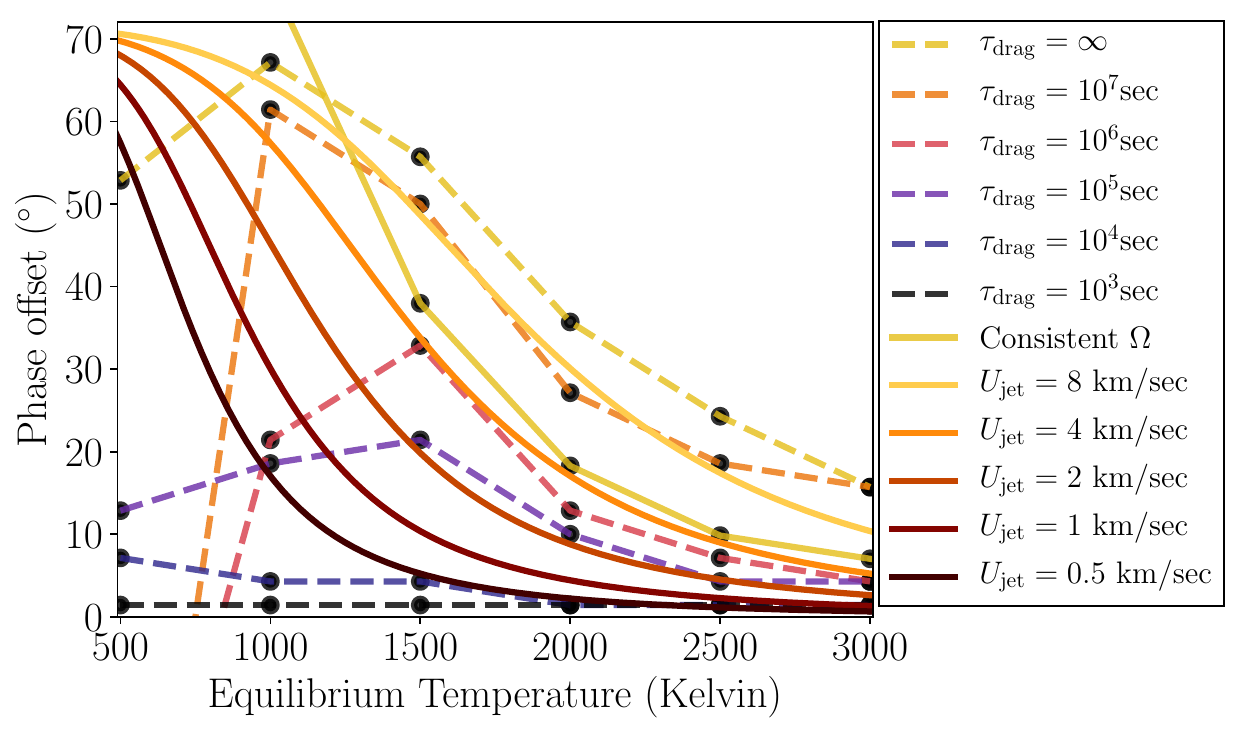}
	\caption{Comparison between numerical (dashed and solid lines, dots) and analytic (solid lines) infrared phase offsets, plotted against global equilibrium temperature. The numerical results are the same as in Figure \ref{fig:phaseshift_numerical}, while the analytic results are plotted using the theory of \cite{Zhang:2016} at $10 \ \mathrm{mbar}$ pressure for varying assumed speeds of the equatorial eastward jet, from $0.5 - 8 \ \mathrm{km/sec}$. Both the analytic theory and numerical results show the same trend of decreasing infrared phase offsets with increasing $T_{\mathrm{eq}}$, as in \cite{Cowan:2011}.}
	\label{fig:phaseshift_numericalvsanalytic}
\end{figure}
\indent \cite{Zhang:2016} developed a simple theoretical model, similar to that of \cite{Cowan:2011}, to determine the phase offset of the planetary hot spot for tidally locked gas giants. To do so, they used a kinematic model in the Newtonian cooling approximation, assuming a horizontally constant radiative timescale and a zonally symmetric jet with a prescribed zonal wind speed. Doing so, they derived the following transcendental relation that determines the hot spot offset for a given equatorial jet speed and radiative timescale (see their Appendix B):
\begin{equation}
\label{eq:lambdamax}
\mathrm{sin}(\lambda_{\mathrm{max}} -\lambda_{\mathrm{s}})e^{\lambda_{\mathrm{max}}/\xi} = -\frac{\eta}{\xi \mathrm{cos}(\lambda_\mathrm{s})} \mathrm{,}
%\lambda_{\mathrm{max}} \approx \mathrm{tan}^{-1}\left(\frac{\tau_{\mathrm{rad}}}{\tau_{\mathrm{adv}}}\right) \mathrm{.}
\end{equation}
where $\lambda_{\mathrm{max}}$ is the hot spot offset, $\xi = \tau_{\mathrm{rad}}/\tau_{\mathrm{adv}}$ is the ratio of radiative to advective timescales where $\tau_{\mathrm{adv}} = a/U_{\mathrm{jet}}$ is the timescale for the equatorial jet (with speed $U_{\mathrm{jet}}$) to advect across a planetary hemisphere, $\lambda_{\mathrm{s}} = \mathrm{tan}^{-1}(\xi)$, and
\begin{equation}
\eta = \frac{\xi}{1+\xi^2}\frac{e^\frac{\pi}{2\xi} + e^{\frac{3\pi}{2\xi}}}{e^{\frac{2\pi}{\xi}}-1} \mathrm{.}
\end{equation}
\cite{Zhang:2016} compared the prediction from \Eq{eq:lambdamax} to their numerical simulations using Newtonian cooling, finding good agreement throughout parameter space (see their Figure 9).   \\
\indent Here we examine the predictions from \Eq{eq:lambdamax} for how phase offset varies with equilibrium temperature in context of our numerical simulations examined in \Sec{sec:numericaloffset}. \Fig{fig:phaseshift_numericalvsanalytic} shows a comparison between the solutions to \Eq{eq:lambdamax} for varying speeds of the equatorial jet and our numerical calculations of infrared phase offsets. For this comparison, we use the relationship between $\tau_{\mathrm{rad}}$ and $T_{\mathrm{eq}}$ from \Eq{eq:taurad} to calculate analytic phase offsets as a function of $T_{\mathrm{eq}}$. \\\ 
\indent The analytic theory of \cite{Zhang:2016} reproduces the general trend of decreasing phase offset with increasing equilibrium temperature from our GCMs with double-grey radiative transfer. However, note that given that there is no theory for the equatorial jet speeds in hot Jupiter atmospheres (which will be discussed further in \Sec{sec:discussion}), the analytic jet speeds do not vary self-consistently with $T_{\mathrm{eq}}$. The phase offsets from our GCMs do not have a monotonic dependence with temperature, due to nonlinear interactions between the equatorial waves driving the superrotating jet\footnote{In two cases, with $T_{\mathrm{eq}} = 500 \ \mathrm{K}$ and $\tau_{\mathrm{drag}} = 10^6 \ \mathrm{and} \ 10^7 \ \mathrm{s}$, the phase shift is westward.}. Additionally, the analytic theory shows a slightly steeper relationship between $\lambda_{\mathrm{max}}$ and $T_{\mathrm{eq}}$ than the GCM calculations with varying $\tau_{\mathrm{drag}}$, and a shallower dependence than our GCM calculations with the rotation rate varying consistently with $T_{\mathrm{eq}}$. This is likely due to the lack of dependence of $U_{\mathrm{jet}}$ on $T_{\mathrm{eq}}$ in the analytic theory.
%and the assumption of a fixed opacity with $T_{\mathrm{eq}}$ in our GCMs. 
Including this would increase the analytic infrared phase offsets at high $T_{\mathrm{eq}}$, as the jet speed should increase with incident stellar flux, decreasing $\tau_{\mathrm{adv}}$. Similarly, the analytic theory appears to predict a much larger phase offset than found in our main grid of GCM calculations (without a consistently varying rotation rate) at low $T_{\mathrm{eq}} = 500 \ \mathrm{K}$. This is mainly because the jet speeds at such a low value of incident stellar flux are very small when drag is applied, with a value of $U_{\mathrm{jet}} \sim 10 \ \mathrm{m} \ \mathrm{s}^{-1}$ for $\tau_{\mathrm{drag}} = 10^6 \ \mathrm{s}$ and $U_{\mathrm{jet}} \lesssim 1  \ \mathrm{m} \ \mathrm{s}^{-1}$ for $\tau_{\mathrm{drag}} \le 10^5 \ \mathrm{s}$.
%The latter has the same effect as a decreased dependence of $\tau_{\mathrm{rad}}$ on $T_{\mathrm{eq}}$, and our numerical infrared phase offsets would be smaller if we used more accurate opacities.
\section{Discussion \& directions for future work}
\label{sec:discussion}
\indent As shown in \Sec{sec:comparisontoobs}, the theory of \cite{Komacek:2015} and \cite{Zhang:2016} agrees well with the observed trend of increasing infrared phase curve amplitude with increasing temperature. Additionally, as shown in \cite{Komacek:2015} this same theory provides a good match to GCMs solving the same set of physical equations, with zero free parameters and no tuning.
%Within the context of the analytic theory, the drag timescale is a free parameter that can be adjusted to increase the theoretical phase curve amplitude to better agree with observations at low equilibrium temperature (Figure \ref{fig:A_theory_individual}). 
However, if drag is stronger than the Coriolis force for a given observed planet, our theory relies on knowledge of the characteristic drag timescale to estimate the phase curve amplitude and hence requires understanding of the underlying processes causing this drag. In our simple theory we utilized a linear (Rayleigh) drag, on both the zonal and meridional components of the wind. Note that small-scale turbulence \citep{Li:2010}, shocks \citep{Heng:2012a}, and Lorentz forces \citep{batygin_2013,Rauscher_2013,Heng:2014,Rogers:2014,Rogers:2020} would, in reality, produce anisotropic drag. The strength and direction of this drag might itself depend on parameters such as incident stellar flux, rotation rate, and atmospheric composition. This necessitates further work using numerical simulations to ascertain how the trend in dayside-nightside temperature differences with equilibrium temperature changes when using a realistic drag formalism.    \\
\indent Though not parameterized in our analytic theory, clouds likely play a large role in affecting the dayside-nightside temperature differences, especially at low equilibrium temperature where we find that our theory systematically under-predicts observed phase amplitudes. Understanding fully the effects of clouds on the apparent dayside-nightside temperature differences in hot Jupiter atmospheres also requires self-consistent numerical simulations of atmospheric circulation including cloud radiative feedbacks on the circulation. Such a model has recently been developed to understand the radiative effects of clouds on the atmospheric circulation of HD 189733b \citep{Lee:2016} and GJ 1214b \citep{Charnay:2015}. However, given that these models are computationally intensive, no thorough self-consistent examination of the effects of clouds in hot Jupiter atmospheres with varying planetary parameters (e.g. incident stellar flux, rotation rate) has been performed. Such a numerical study would aid in determining whether or not the discrepancies between our theory and observations at relatively low incident stellar flux are due to nightside cloud decks reducing the outgoing infrared flux. \\
%discuss non-grey effects on theoretical predictions \\
\indent To date, there is no developed theory for what controls the speed of the eastward equatorial jet in hot Jupiter atmospheres, though its qualitative formation mechanism is well understood \citep{Showman_Polvani_2011,Tsai:2014}. Notably, this prevents the development of a self-consistent predictive formalism for infrared hot spot offsets. In \Sec{sec:compareoffset} we compared our simulations to the theory of \cite{Zhang:2016}, which requires the speed of the equatorial jet as an input parameter. In \cite{Zhang:2016}, the jet speed was taken from their numerical simulations, and in this work we have simply taken it to be a free parameter. However, linking theoretically the phase offset to the dayside-nightside temperature differences using a fluid dynamical approach would be a useful improvement on the kinematic theory of \cite{Cowan:2011} and \cite{Zhang:2016}. Such a theory could in principle be developed, as the same equatorial wave action that drives the equatorial jet acts to reduce dayside-nightside temperature differences. This theory would allow for consistent prediction of phase offsets and dayside-nightside temperature differences, and, if the underlying drag mechanisms are better understood, could in principle allow complete first-order understanding of the general circulation of tidally locked gas giant atmospheres from phase curve observations.   
\section{Conclusions}
\label{sec:conclusions}
\begin{enumerate}
\item The theory of \cite{Komacek:2015} predicts that fractional dayside-nightside temperature differences in hot Jupiter atmospheres increase with increasing planetary equilibrium temperature. Infrared phase curve observations also exhibit such a trend, and in this work we showed that the theory agrees reasonably well with the observed trend. When applied to individual planets, this theory matches well at high equilibrium temperatures $\gtrsim 2000 \ \mathrm{K}$. However, for an assumed fixed photosphere pressure of $100 \ \mathrm{mbar}$ and assuming no drag, it systematically under-predicts the dayside-nightside temperature differences for planets with equilibrium temperatures $\lesssim 2000 \ \mathrm{K}$. This is likely due to a process that decreases the pressure at which the photosphere lies for cooler hot Jupiters, which could potentially be supersolar metallicities enhancing the atmospheric opacity or clouds muting the emitted flux from the nightside of the planet. 
%This is most likely due to clouds muting the infrared emission from the nightside of these less strongly irradiated hot Jupiters, increasing the amplitude of observed phase curves. 
\item A suite of numerical simulations including double-grey radiative transfer shows qualitatively similar trends of increasing dayside-nightside temperature differences with increasing equilibrium temperature and drag strength. The analytic theory correctly predicts that the transition between low and high dayside-nightside temperature differences with varying drag strengths in our numerical simulations occurs between $\tau_{\mathrm{drag}} = 10^5 \ \mathrm{s}$ and $\tau_{\mathrm{drag}} = 10^4 \ \mathrm{s}$. However, due to the use of opacities that do not vary with incident stellar flux, the numerical simulations show a shallower trend of increasing dayside-nightside temperature differences with increasing incident stellar flux than both the observations and our analytic predictions.
%Post-processing these numerical simulations, we find that without including chemical effects the phase amplitude at $3.6 \ \mu \mathrm{m}$ should be much lower than that at $4.5 \ \mu \mathrm{m}$, opposite to that seen in infrared phase curves of HD 189733b and WASP-14b. This potentially points to disequilibrium chemical effects increasing the abundance of methane on the nightside of these planets, decreasing the amount of nightside outgoing infrared flux and raising the phase curve amplitude.  
\item Both our numerical simulations with double-grey radiative transfer and the analytic theory of \cite{Zhang:2016} predict similar decreases in eastward infrared phase offset with increasing incident stellar flux. Using varied values of drag timescales (or equatorial jet speeds), these models can explain almost the entire set of observed infrared phase offsets. However, there is some degree of scatter in these observations, and as a result it is yet to be seen whether there is a clear trend in observed phase offset with incident stellar flux. 
\end{enumerate}
\acknowledgements
We thank Jacob Bean, Ian Crossfield, Sivan Ginzburg, Cheng Li, Mike Line, Vivien Parmentier, Everett Schlawin, Maria Steinr\"{u}ck, Rob Zellem, Xi Zhang, and the Steward Observatory Planet Theory group for insightful discussions. We thank Kevin Heng for a thorough review, which improved the manuscript. This research was supported by NASA Origins grant NNX12AI79G to A.P.S. T.D.K. acknowledges support from NASA headquarters under the NASA Earth and Space Science Fellowship Program Grant PLANET14F-0038.  X.T. also acknowledges support from NASA headquarters under the NASA Earth and Space Science Fellowship Program (ASTRO). Resources supporting this work were provided by the NASA High-End Computing (HEC) Program through the NASA Advanced Supercomputing (NAS) Division at Ames Research Center. This research has benefited from the Opacity Wizard tool developed by Caroline Morley.

\appendix

\section{Demonstration of the applicability of {\twostr} to hot Jupiter atmospheres}
\label{appendix}
\begin{figure*}
	\centering
	\includegraphics[width=0.75\textwidth]{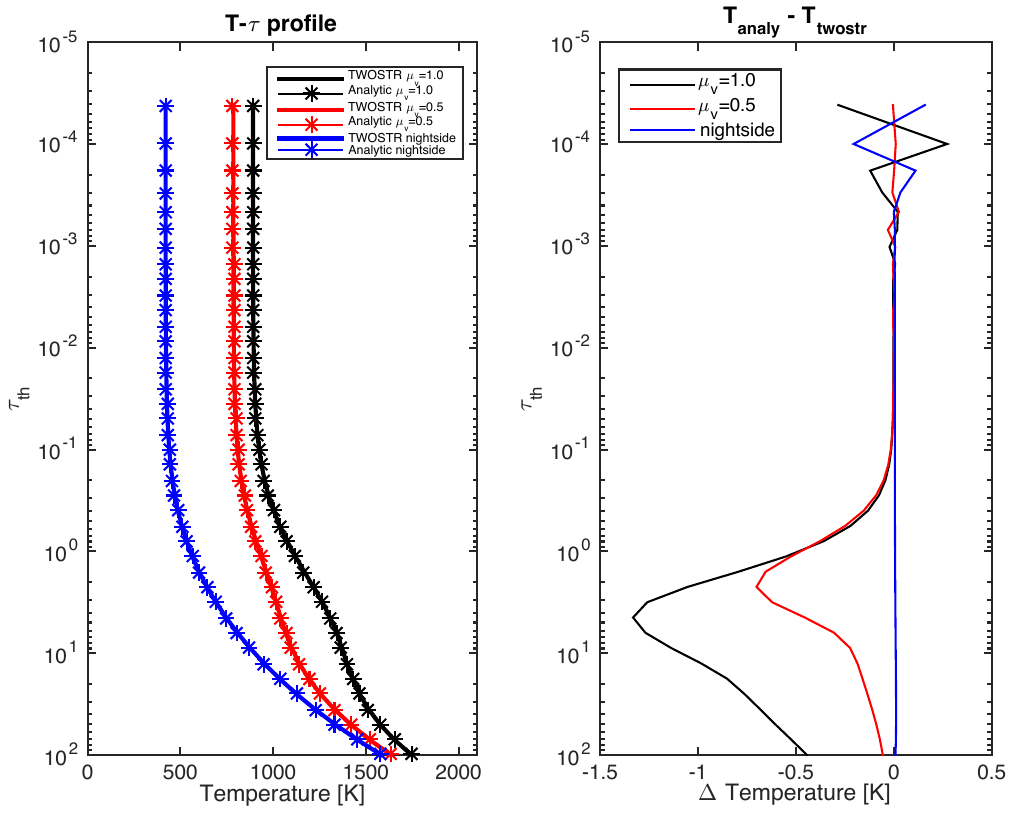}
	\caption{Comparison between {\twostr} solutions for the $T-\tau$ profile and analytic solution of Equation (\ref{eq:guillot}) (left) and temperature difference between analytic and numerical calculations (right) for varying stellar zenith angles $\mu_v$. For this comparison, we set the irradiation temperature $T_{\mathrm{irr}} = 1000 \ \mathrm{K}$, internal temperature $T_{\mathrm{int}} = 500 \ \mathrm{K}$, thermal opacity $\kappa_{\mathrm{th}} = 10^{-3} \mathrm{m}^2 \ \mathrm{kg}^{-1}$, visible opacity $\kappa_{\mathrm{v}} = 4 \times 10^{-4} \mathrm{m}^2 \ \mathrm{kg}^{-1}$, and thermal zenith angle $\mu_{\mathrm{th}} = 0.5$. We find a maximum discrepancy less than $1.5 \ \mathrm{K}$ down to an infrared optical depth $\tau_{\mathrm{th}} = 100$, demonstrating the accuracy of the {\twostr} calculation.}
	\label{fig:rttest}
\end{figure*}
To demonstrate that the {\twostr} package is correctly implemented in the {\mitgcm} and its applicability in the hot Jupiter regime, we compare a numerical calculation of the radiative equilibrium temperature profile as a function of optical depth to an analytic solution, as in \cite{Rauscher_2012}. To find an analytic solution for the radiative equilibrium temperature structure, first, manipulate Equations (\ref{rteq1}) and (\ref{rteq2}) using $I = I^+ - I^-$ and $\overline{I} = (I^+ + I^-)/2$, and set $q=0$ in Equation (\ref{heatingrate}) at all optical depths. Applying boundary conditions of visible flux $F_0 = \sigma T_{\mathrm{irr}}^4$ at the top of the atmosphere and net upward internal thermal flux $F_{\mathrm{int}} = \sigma T_{\mathrm{int}}^4$ at the bottom, we find the radiative equilibrium temperature profile
\begin{equation}
\label{eq:guillot}
\begin{aligned}
T^4(\tau)  = & \frac{T^4_{\mathrm{irr}} \mu_\mathrm{v}}{2}\left[1 + \alpha + \left(\frac{1}{\alpha} - \alpha\right)e^{-\tau/(\mu_\mathrm{v}\gamma)}\right]  \\
& + \frac{T^4_{\mathrm{int}}}{2}\left(1+\frac{\tau}{\mu_{\mathrm{th}}}\right) \mathrm{.}
\end{aligned}
\end{equation}
In \Eq{eq:guillot}, $\gamma = \kappa_{\mathrm{th}}/\kappa_{\mathrm{v}}$, where $\kappa_{\mathrm{th}}$ and $\kappa_{\mathrm{v}}$ are the absorption coefficient of the thermal and visible bands, respectively, $\alpha = \gamma \mu_{\mathrm{v}}/\mu_{\mathrm{th}}$, where the visible zenith angle $\mu_{\mathrm{v}} = \mathrm{cos} \left(\theta_{\mathrm{v}}\right)$ and infrared zenith angle $\mu_{\mathrm{th}} = \mathrm{cos} \left(\theta_{\mathrm{th}}\right)$. We set $\mu_{\mathrm{th}} = 0.5$, as in \cite{Kylling:1995}. This solution is almost identical to the two-stream solution of \cite{Guillot:2010}, except here we use a different $\mu_{\mathrm{th}}$ to be consistent with the initial implementation of {\twostr}. We show a comparison between the $T-\tau$ profile of {\twostr} and \Eq{eq:guillot} in \Fig{fig:rttest}. There is nearly perfect agreement between {\twostr} calculated and analytic solutions at $\tau_{\mathrm{th}} < 0.1$, and $\lesssim 1.5 \ \mathrm{K}$ discrepancy at greater optical depths, demonstrating the excellent applicability of our double-gray tool to hot Jupiters.
\if\bibinc n
\bibliography{References}
\fi

\if\bibinc y

\fi

\end{document}